\documentclass[twocolumn,aps,showpacs,prl,superscriptaddress, nourl]{revtex4-1}
\usepackage[latin9]{inputenc}
\usepackage{verbatim}
\usepackage{amsmath}
\usepackage{amssymb}
\usepackage{graphicx}
\usepackage{esint}
\usepackage{hyperref}
\usepackage{xcolor}
\usepackage{soul}
\makeatletter
\@ifundefined{textcolor}{}
{%
 \definecolor{BLACK}{gray}{0}
 \definecolor{WHITE}{gray}{1}
 \definecolor{RED}{rgb}{1,0,0}
 \definecolor{GREEN}{rgb}{0,1,0}
 \definecolor{BLUE}{rgb}{0,0,1}
 \definecolor{CYAN}{cmyk}{1,0,0,0}
 \definecolor{MAGENTA}{cmyk}{0,1,0,0}
 \definecolor{YELLOW}{cmyk}{0,0,1,0}
}

\newcommand{\Fig}[1]{Fig.~\ref{#1}}

\makeatother

\begin{document}

\title{Spectral evolution of the SU(4) Kondo effect from the single impurity to
the two-dimensional lattice}
\author{Alejandro M. Lobos}
\email{alobos@umd.edu}
\affiliation{Joint Quantum Institute and Condensed Matter Theory Center, Department of
Physics, University of Maryland, College Park, Maryland 20742, USA}
\author{Marcelo Romero}
\affiliation{Instituto de Desarrollo Tecnol{\'{o}}gico para la Industria Qu{\'{\i}}mica
(INTEC-CONICET-UNL) G{\"{u}}emes 3450, CC91, (S3000GLN), Santa F{\'{e}},
Argentina.}
\author{Armando A. Aligia}
\affiliation{Centro At{\'{o}}mico Bariloche and Instituto Balseiro, Comisi{\'{o}}n
Nacional de Energ{\'{\i}}a At{\'{o}}mica, 8400 Bariloche, Argentina.}
\date{\today}

\begin{abstract}
We describe the evolution of the SU(4) Kondo effect as the number of magnetic centers increases from one impurity to the two-dimensional (2D) lattice. 
We derive a Hubbard-Anderson model which describes a 2D array
of atoms or molecules with two-fold orbital degeneracy, acting as magnetic impurities and interacting with a metallic host. We calculate the differential conductance, observed typically in experiments of scanning tunneling spectroscopy, for different arrangements 
of impurities on a metallic surface: a single impurity, a periodic square lattice,
and several sites of a rectangular cluster. Our results point towards the crucial importance of the orbital degeneracy and agree well with recent experiments in different systems of iron(II) phtalocyanine molecules deposited on top of Au(111) [N. Tsukahara \textit{et al.}, Phys. Rev. Lett. \textbf{106}, 187201 (2011)], indicating  that this would be the first experimental realization of an artificial 2D SU(4) Kondo-lattice system.
\end{abstract}


\date{\today }
\maketitle

The Kondo effect is one of the most paradigmatic phenomena in strongly
correlated condensed matter systems \cite{hewson}. 
It is characterized by the emergence of a many-body singlet ground state formed by
the impurity spin and the conduction electrons in the Fermi sea, which form a screening 
{\textquotedblleft}cloud{\textquotedblright} around the impurity.
Originally observed in dilute magnetic alloys \cite{hewson}, the Kondo
effect has reappeared more recently in the context of semiconductor
quantum-dot (QD) systems \cite{cronenwet98,goldhaber98}, and in systems of
magnetic adatoms (e.g., Co or Mn) deposited on clean metallic surfaces,
where the effect has been clearly observed experimentally as a narrow Fano-Kondo 
antiresonance (FKA) in the differential conductance in scanning tunneling
spectroscopy (STS) \cite{li98,Madhavan98_Tunneling_into_single_Kondo_adatom,knorr02}. 

While most of the experimental realizations of the Kondo effect correspond to spin
1/2 and SU(2) symmetry, more exotic Kondo effects are possible in nanoscopic systems \cite{Galpin05_QPT_in_double_QDs, *Mitchell10_2CK_in_triple_QDs, *DiNapoli13_NFL_in_Co_chains, *Kuzmenko13_2CK_in_double_QDs_with_SO(n)_symmetry}.
In particular, a SU(4) Kondo effect can occur when an additional pseudospin 1/2 orbital degree of 
freedom appears due to robust orbital degeneracy.
In practice, however, the stringent conditions to preserve orbital degeneracy limits the observation of the SU(4) Kondo effect to few cases, such as
C nanotubes \cite{jarillo05,busser07,anders08}, and Si fin-type field effect transistors \cite{tetta12} where there is a valley degeneracy \cite{roura12}.
Recently, Minamitani \textit{et al.} \cite{Minamitani12_SU4_Kondo_in_FePc_molecules} have shown that the Kondo effect observed in isolated iron(II) phtalocyanine (FePc) molecules deposited on top of clean Au(111) (in the most usual
on-top configuration) \cite{Gao07_Site_specific_Kondo_effect} is a new realization of the SU(4) case. In the on-top configuration, the degeneracy between partially filled $3d_{xz}$ and $3d_{yz}$ orbitals of Fe is preserved by the Au(111) substrate, leading to a strong FKA in the STS signal.
Interestingly, Tsukahara \textit{et al. } \cite{Tsukahara10_Evolution_of_Kondo_resonance} showed  that at sufficiently high densities, the  FePc molecules on Au(111) self-organize into a two-dimensional (2D) square lattice, 
paving the way to study artificially engineered Kondo lattices by scanning tunneling microscopy (STM).
At present, a large class of organic-Kondo adsorbates are being studied by STM techniques due to their potential 
applications as electronic  \cite{venka06,wang10} 
and/or molecular spintronics \cite{iancu06,perera10,garnica13} devices,
and therefore it is important to understand their electronic properties. 
Recent \textit{ab-initio} calculations have demonstrated the crucial role of the interaction
between the organometallic molecule and the substrate for designing spintronic devices \cite{Mugarza12_Metal_Pc_on_Ag100, *Gargiani13_Metal_Pc_molecules_on_Au110}.
In this context, the effect of the orbital degrees of freedom in artificially engineered Kondo lattice systems 
remains to be explored, and to the best of our knowledge the extension of the SU(4) impurity model to the lattice has not been studied so far.

Motivated by these recent developments, in this Letter, we theoretically study the
evolution of the SU(4) 
Kondo effect, from the single impurity to
the 2D Kondo-lattice limit. 
Guided by general symmetry principles, we derive an effective SU(4) Hubbard-Anderson model describing 
coupled magnetic impurities with an additional orbital degree of freedom, forming clusters on the metallic substrate. While our results are generic, and in principle  applicable to other organometallic Kondo systems, in what follows we specify our results for the case of Ref. \onlinecite{Tsukahara10_Evolution_of_Kondo_resonance}, as we believe this to be the first realization of an artificial 2D SU(4) Kondo lattice. 
We calculate the STS differential conductance $dI/dV$ (as observed
experimentally), and analyze the line shapes upon variation of the size and connectivity of the cluster. 
Our results show a good agreement with
experiment and are important for the correct physical interpretation of the data. 
In particular, we show that the most prominent feature of the experiment (i.e., the splitting of the FKA in the case of 
high coordination number \cite{Tsukahara10_Evolution_of_Kondo_resonance})
is a consequence of the orbital degeneracy \cite{note}.
As explained below, this opens new exciting possibilities, 
such as the existence of new phases with orbitally-ordered ground states 
\cite{Khomskii73_Orbital_and_magnetic_ordering_in_2D, Aligia04_Magnetic_and_orbital_order_in_ruthenates}.

\begin{figure}
\includegraphics[scale=0.32, viewport=10 70 700 450,clip]{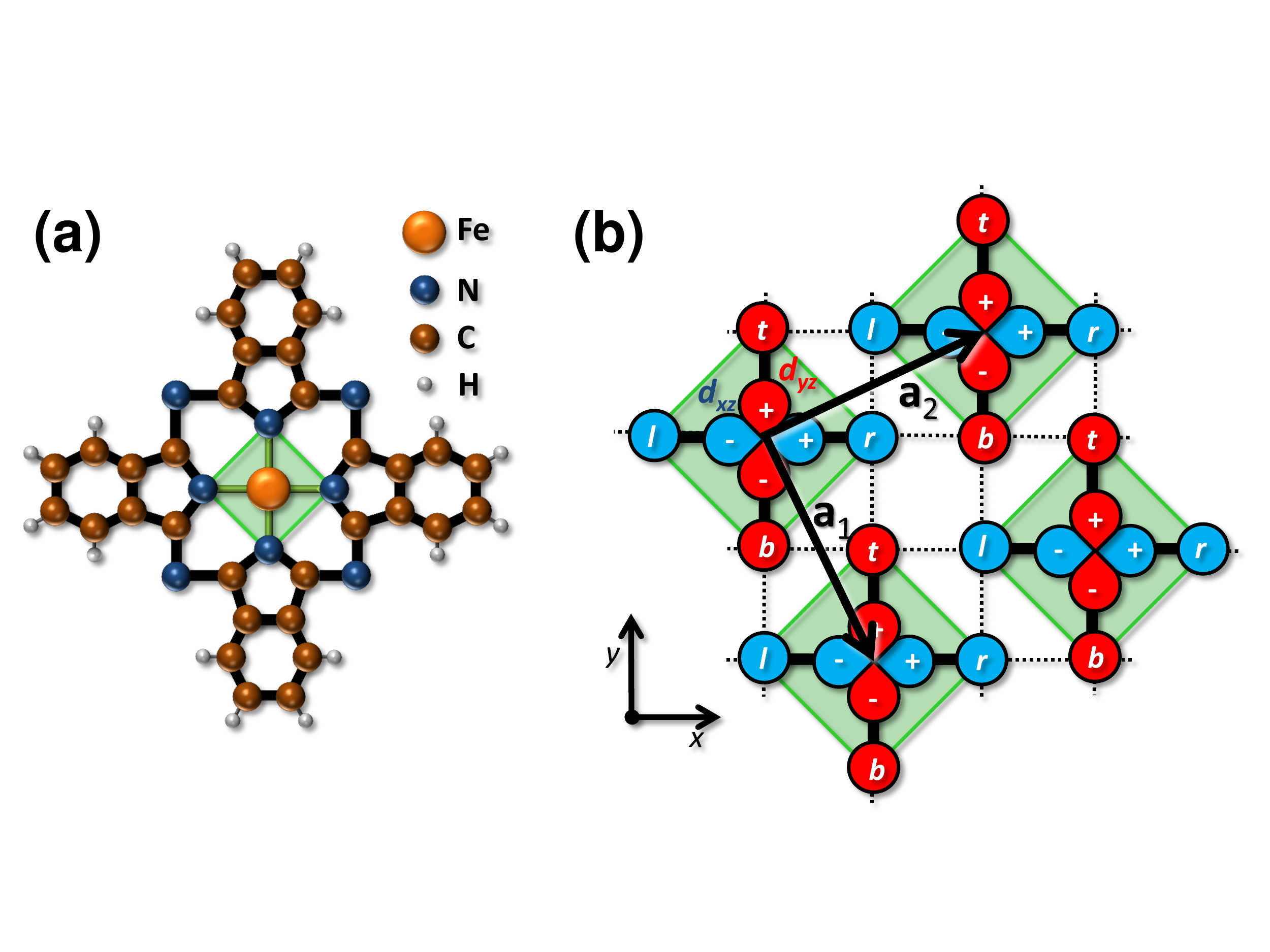}
\caption{(Color online) 
(a) Representation of a iron(II) phtalocyanine (FePc) molecule. 
The region shaded in green is the FeN$_4$ substructure which is kept in the theoretical model. 
(b) System of  FeN$_4$ molecules forming a cluster.} 
\label{fig:system}
\end{figure}

\textit{Model.-} We 
derive an
effective minimal Hubbard-Anderson model for the 2D lattice of FePc
molecules. For the case of an isolated molecule (see Fig. \ref{fig:system}(a)), the effective SU(4)
Anderson model has been derived previously
\cite{Minamitani12_SU4_Kondo_in_FePc_molecules}. 
The low-energy physics is
described by two degenerate molecular orbitals of $xz$ and $yz$ symmetry,
which have most of their weight on the corresponding $3d$ orbitals of the Fe
atom.
To extend this impurity model to the lattice, we add the hopping
between nearest-neighbor (NN) molecules, leading to a model similar to the
one used to describe a trimer of Co atoms on Au(111) \cite{aaa06}. However,
in the present case, the orbital degeneracy and the symmetry of the
molecular orbitals introduce peculiar features. On general
symmetry grounds, one expects that the effective hopping between any two NN
molecular orbitals 
will depend on the \textit{direction} of the hopping.
In particular, we assume that the effective hopping between NN $3d$ Fe orbitals can occur 
either by direct overlap of the organic ligands, or via the Au substrate.
In the first case, the coupling can be thought as occurring via  
the $p_z$ orbitals of the neighboring N atoms. 
Defining the $x$ and $y$ directions as those pointing from the Fe atom to the organic ligands in the molecule,
as in Fig. 1, the Fe $3d_{\nu z}$ hybridizes only with the $p_z$ orbitals of the N atoms in 
the $\nu$ direction ($\nu$ = $x$ or $y$), and the hopping with other orbitals vanishes by symmetry. 
The presence of the substrate modifies these arguments (see Appendix), but 
the crucial directional dependence of the effective hopping is a robust feature that remains.

The effective model is $H=H_{mol}+H_{c}+H_{mix}$,
where $H_{mol}$ describes the molecular states and the hopping between them, 
$H_{c}$ the conduction states, and $H_{mix}$ the coupling between them.
To illustrate the derivation of $H_{mol}$, we have calculated the effective
hopping between molecular orbitals in a lattice of hypothetical FeN$_{4}$
molecules (i.e., the central part of FePc) as shown in Fig. \ref{fig:system}. For
each molecule, the relevant molecular states are:
\begin{align}
\left\vert \tilde{x}_{\mathbf{r}_{ij},\sigma }\right\rangle & =\left[ \alpha 
\tilde{d}_{\mathbf{r}_{ij},\sigma }^{x}+\beta \left( \tilde{p}_{\mathbf{r}%
_{ij},\sigma }^{\left( r\right) }-\tilde{p}_{\mathbf{r}_{ij},\sigma
}^{\left( l\right) }\right) \right] ^{\dagger }\left\vert 0\right\rangle , 
\notag \\
\left\vert \tilde{y}_{\mathbf{r}_{ij},\sigma }\right\rangle & =\left[ \alpha 
\tilde{d}_{\mathbf{r}_{ij},\sigma }^{y}+\beta \left( \tilde{p}_{\mathbf{r}%
_{ij},\sigma }^{\left( t\right) }-\tilde{p}_{\mathbf{r}_{ij},\sigma
}^{\left( b\right) }\right) \right] ^{\dagger }\left\vert 0\right\rangle .
\label{xy}
\end{align}%
Here, $\tilde{d}_{\mathbf{r}_{ij},\sigma }^{\nu}$ is the destruction operator for electrons with
spin $\sigma $ in the $3d_{\nu z}$  orbital of Fe at cluster with
position $\mathbf{r}_{ij}=i\mathbf{a}_{1}+j\mathbf{a}_{2}$ 
(with $\mathbf{a}_{1},\mathbf{a}_{2}$ the Bravais lattice vectors defined in Fig. \ref%
{fig:system}(b)), and $\tilde{p}_{\mathbf{r}_{ij},\sigma }^{\left( \eta
\right) }$ is the destruction operator in the 2$p_{z}$ orbital of the N atom
located at position $\eta =\left\{ r,l,t,b\right\} $ within the molecule
(respectively: right, left, top, bottom, with respect to the central Fe atom
in the molecule).

It is easy to calculate the effective hopping between molecular states, in a
tight-binding description, assuming a hopping $t^\prime$ between NN N atoms (see dotted lines in Fig. \ref{fig:system}(b))
(see Appendix). The magnitude of this hopping is
either $t=\left|\beta\right|^{2}t^\prime$ or zero. To simplify the model, one
can ``rotate'' the molecular orbitals defining a new basis set 
$\left\{ \left|x_{\mathbf{r}_{ij},\sigma}\right\rangle ,\left|y_{\mathbf{r}%
_{ij},\sigma}\right\rangle \right\} $ such that $\left\langle x_{\mathbf{r}%
_{ij},\sigma}\right|H\left|y_{\mathbf{r}_{lm},\sigma}\right\rangle =0$, for
all $\mathbf{r}_{ij},\mathbf{r}_{lm}$, therefore conserving the orbital index 
$\nu=\left(x,y\right)$ in the hopping process.
It is more convenient for us to work in the hole representation.
Calling $h_{%
\mathbf{r}_{ij},\sigma}^{\nu}$ the operators which destroy a hole 
(create an electron) in the
molecular state $\left|\nu_{\mathbf{r}_{ij},\sigma}\right\rangle$ in the new basis (see Appendix), we arrive at the effective 2D Hubbard model:
\begin{align}
H_{mol}& =\sum_{ij}^{N}\left[ -\sum_{\sigma ,\nu }\left( t_{2}h_{\mathbf{r}%
_{ij},\sigma }^{\nu \dagger }h_{\mathbf{r}_{ij}\pm \mathbf{a}_{\nu },\sigma
}^{\nu }+t_{1}h_{\mathbf{r}_{ij},\sigma }^{\bar{\nu}\dagger }
h_{\mathbf{r}_{ij}\pm \mathbf{a}_{\nu },\sigma }^{\bar{\nu}}\right) \right.   \notag \\
& \left. + E_{h}n_{\mathbf{r}_{ij}}+\frac{U}{2}n_{\mathbf{r}_{ij}}
\left( n_{\mathbf{r}_{ij}}-1\right) \right],  
\label{hmol}
\end{align}
where 
the effective hopping amplitudes $t_{1}$ and $t_{2}$ connect NN 
$h^{\nu }$ orbitals located at $\mathbf{r}_{ij}$ and $\mathbf{r}_{ij}\pm 
\mathbf{a}_{\nu }$, with the compact notation 
$\left( \mathbf{a}_{x}=\mathbf{a}_{1},\ \mathbf{a}_{y}=\mathbf{a}_{2}\right) $, 
and $\left( \bar{x}=y,\ 
\bar{y}=x\right) $. $E_{h}$ and $n_{\mathbf{r}_{ij}}=\sum_{\sigma \nu }n_{\mathbf{r}_{ij},\sigma }^{\nu }$, with
$n_{\mathbf{r}_{ij},\sigma }^{\nu }=h_{\mathbf{r}_{ij},\sigma }^{\nu \dagger }h_{\mathbf{r}_{ij},\sigma }^{\nu }$
are, respectively, the energy and number of holes. The last term in
Eq. (\ref{hmol}) accounts for the local Hubbard repulsion between holes at
site $\mathbf{r}_{ij}$. Note that Hamiltonian Eq. (\ref{hmol}) is explicitly
SU(4)-invariant. For the simplified system of FeN$_{4}$ molecules we obtain $%
t_{1}=0.618t$ and $t_{2}=-1.618t$. In the case of an effective hopping mediated by conduction states in the substrate we obtain the same qualitative features: it is highly anisotropic and conserves the orbital index (see Appendix).

To consider the coupling to the metallic substrate, we assume that the
distance between the Hubbard sites is $R\gg1/k_{F}$, with $k_{F}$ the Fermi
momentum of the metallic substrate (see Appendix). This approximation is not generic, but
this limit is well verified in experimental molecular Kondo systems, 
and permits to
neglect indirect correlations among Hubbard sites mediated by the metal
[such as  Ruderman-Kittel-Kasuya-Yosida (RKKY) interactions or coherent Kondo correlations arising from the
overlap of Kondo screening clouds] 
\cite{aaa06,Romero11_STM_for_adsorbed_molecules,Barzykin00_Kondo_cloud,Simonin07_Kondo_cloud,Affleck10_Kondo_cloud,Lobos12_Dissipative_XY_chain, Lobos13_FMchains}. In such a limit, the 2D metal can be effectively described by a collection
of uncorrelated ``fermionic baths'', each one coupled to each Hubbard site $\mathbf{r}_{ij}$ 
\cite{Lobos12_Dissipative_XY_chain, Lobos13_FMchains}.
Therefore, we describe the metallic substrate as 
$H_{c} =\sum_{i j \xi \sigma \nu }\epsilon _{\xi}c_{\mathbf{r}_{ij},\xi,\sigma }^{\nu \dagger }c_{\mathbf{r}_{ij},%
\xi,\sigma }^{\nu }$, 
where $c_{\mathbf{r}_{ij},\xi,\sigma}$ is the annihilation operator of a conduction hole with spin $\sigma$ and quantum number $\xi$ at position $\mathbf{r}_{ij}$. 
The coupling to the molecules is described by
$H_{mix} =V
\sum_{i j \xi \sigma \nu }\left( h_{\mathbf{r}_{ij},\sigma }^{\nu \dagger }c_{\mathbf{r}_{ij},\xi,\sigma
}^{\nu }+\text{H.c.}\right) $ (see Appendix).

We note that $H$ is a many-body Hamiltonian
which cannot be solved exactly.
Assuming the limit of strong repulsion $U\rightarrow \infty$, 
we can neglect configurations with two or more holes in a molecular orbital, 
and consider only local charge fluctuations between the subspaces with $n=0,1$ holes. 
This limit can be implemented in 
the slave-boson representation \cite{coleman84,*Coleman87,*newns87} $h_{\mathbf{r}%
_{ij},\sigma }^{\nu }=b_{\mathbf{r}_{ij}}^{\dagger }f_{\mathbf{r}%
_{ij},\sigma }^{\nu }$, where $b_{\mathbf{r}_{ij}}$ is a bosonic variable
describing the $n_{h}=0$ state (both molecular levels occupied with both
spins) and $f_{\mathbf{r}_{ij},\sigma }^{\nu }$ is a renormalized hole
operator. 
These operators must be constrained by the relation $b_{\mathbf{r}%
_{ij}}^{\dagger }b_{\mathbf{r}_{ij}}+\sum_{\sigma ,\nu }f_{\mathbf{r}%
_{ij},\sigma }^{\nu \dagger }f_{\mathbf{r}_{ij},\sigma }^{\nu }=1.$ 
This
representation of SU($\mathcal{N}$)-invariant Kondo impurities is
particularly useful for $\mathcal{N}\rightarrow \infty $, where the
saddle-point slave-boson mean field approximation
(SBMFA) for the bosonic degrees of freedom 
$b_{\mathbf{r}_{ij}}=b_{\mathbf{r}_{ij}}^{\dagger }
=\left\langle b_{\mathbf{r}_{ij}}\right\rangle =z$ 
becomes exact \cite{coleman84,Coleman87,newns87}. 
After the SBMFA (obtained by replacing 
$h_{\mathbf{r}_{ij},\sigma }^{\nu }\rightarrow 
z f_{\mathbf{r}_{ij},\sigma }^{\nu }$) 
$H$ becomes exactly solvable, and we set $\mathcal{N}=4$ (see Appendix). Physically, the SBMFA describes non-interacting Fermi quasiparticles with renormalized mass $m_e^*/m_e \approx 1/z^2$ and quasiparticle weight $z^2$ near the Fermi level \cite{coleman84,*Coleman87,*newns87}, providing a correct  description of the Kondo-lattice near the Fermi-liquid fixed point.

In STM experiments, the relevant observable is the differential conductance 
$dI/dV$, which in the limit of weak tunneling coupling between the STM tip
and the system becomes proportional to the spectral density 
$dI/dV \sim \rho _{t}\left( -eV\right)$, where the minus sign is needed to pass 
from hole to electron representation, and $t$ represents 
a mixed operator $t_{\mathbf{r}_{ij},\sigma }^{\nu } = 
 \sum_{\xi}c_{\mathbf{r}_{ij},\xi,\sigma }^{\nu }+qh_{\mathbf{r}_{ij},\sigma }^{\nu },$ 
 (with $q$ the Fano parameter), reflecting the
interference between molecule and substrate states  as sensed by the STM tip 
\cite{al05,Romero11_STM_for_adsorbed_molecules,figgins10}. 
We calculate the density of $t$-states as 
$\rho _{t}\left( \omega \right)  =-\frac{1}{\pi }\sum_{\sigma ,\nu }\text{%
Im }\left[ G_{\mathbf{r}_{ij},\nu ,\sigma }^{tt}\left( \omega +i0^{+}\right) \right]$, 
with $G_{\mathbf{r}_{ij},\nu ,\sigma }^{tt}\left( \omega +i0^{+}\right)$ 
the retarded local Green's function of the operator $t_{\mathbf{r}_{ij},\sigma }^{\nu }$.

\textit{Results.-} 
We assume a constant density of conduction states
$\rho=$ 0.137/eV per spin, extending from $-W$ to $W=3.65$ eV. These values are similar to those that provide 
a good fit of the observed line shape for a Co impurity on Cu(111) \cite{al05}. The energy of the molecular 
states (in the hole representation) $E_h$ was taken near to -0.1 eV, according to \textit{ab-initio} 
calculations which find spectral density of Fe $3d_{xz}$ and $3d_{yz}$ states 0.1 eV above the Fermi energy
\cite{Minamitani12_SU4_Kondo_in_FePc_molecules}. We also keep the ratio of hoppings $t_2/t_1$=-3, similar to the values 
obtained above for the simplified system (good fits are also obtained for other values). 
$t_1$,  $V$ and  $q$ are taken as fitting parameters. We define $\Gamma= \pi \rho V^2$.
\begin{figure}[tbp]
\includegraphics[scale=0.3, viewport= 90 40 670 530, clip]{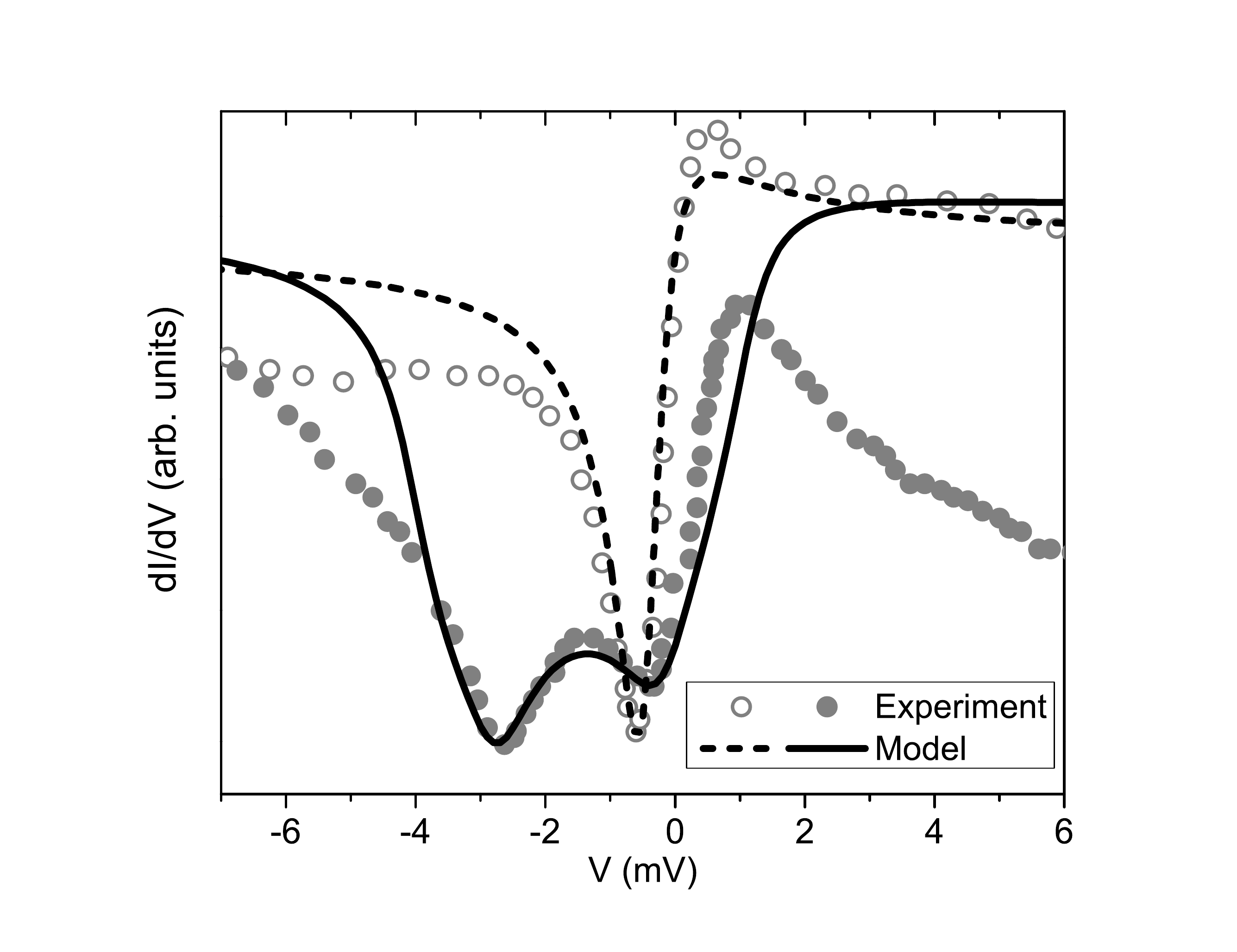}
\caption{Differential conductance as a function of voltage
for an isolated molecule (open circles and dashed line) and the 2D lattice
(solid circles and full line). Circles corresponds to experiment 
\cite{Tsukahara10_Evolution_of_Kondo_resonance} 
and lines to theory with $\Gamma =10.12$ meV. For one molecule $E_h=-112$ meV,  and $q=-0.025$.
For the lattice $E_h=-128$ meV, $t_1=7$ meV, $t_2=3t_1$  and $q=-0.006$.}
\label{1yred}
\end{figure}

In \Fig{1yred} we display our fits of the observed $dI/dV$. From the \textit{ab-initio} 
calculations \cite{Minamitani12_SU4_Kondo_in_FePc_molecules} one can estimate $U=1.6$ eV, which turns out to be 
much larger than the value $\Gamma \approx 0.01$ eV that results from the fit of the isolated molecule. 
It is also much larger 
than $t_i$.
Therefore, the limit 
$U\rightarrow \infty$ 
is well justified. 
Fig. \ref{1yred} shows a good agreement between our theoretical results and the experiment, 
in accordance  with previous results on single Co impurities on Cu(111) \cite{al05}. However, 
in contrast to that case, here the experimental curves had to be slightly shifted 0.55 mV to the left 
to make both curves coincide. This might be related with
experimental uncertainties \cite{noteshift}.

The situation is more difficult for the case of the lattice, because of the double dip structure 
of the observed FKA. 
We have kept the same $\Gamma$ obtained from the fit of the single molecule, but we had to increase slightly  
the magnitude of $E_h$ to $|E_h|=0.128$ eV in order to obtain better fits. This is well justified by the fact that the molecular states, and in particular 
the Fe $3d$ orbitals, increase their occupancy when the molecule is adsorbed on the Au surface \cite{Minamitani12_SU4_Kondo_in_FePc_molecules}, 
and the single-electron levels are expected to increase their energy due to interatomic Coulomb repulsion. 
In addition, we had to modify slightly the value of $q$ to $q=-0.006$, a fact that might be related to 
the different experimental conditions in which the single molecule and lattice $dI/dV$ spectra were obtained in 
Ref. \onlinecite{Tsukahara10_Evolution_of_Kondo_resonance}. 
As shown in Fig. \ref{1yred}, our theory is able to provide semi-quantitative agreement with the experiment. 
In particular, note that the shape of the experimental curve near $V=0$ is well reproduced. As before, 
we have shifted the experimental curve to the left by $1.1$ mV. 

The double-dip structure is a consequence of correlation effects combined with 
the van Hove singularities (VHS) in the spectral density of $H_{mol}$, directly related to the different $|t_1|\neq |t_2|$ in Eq. (\ref{hmol}) [see Ref. 
\onlinecite{notevhs}]. In the SBMFA, the splitting of VHS is given by $\Delta = 4z^2||t_1|-|t_2||$, 
where  the quasiparticle weight $z^2$  introduces a band-narrowing effect due to correlations. In the case of Fig. \ref{1yred} (solid lines), the minimum of the ground state energy is obtained for $z^2 \approx 0.045$, which results in a splitting of $\Delta \approx 2.5$ meV, consistent with the experimentally observed one. 
This value of $z^2$ points to a strongly renormalization effect near the Fermi surface, with a  mass enhancement $m_e^*/m_e \approx 20$.
Correlations are therefore essential to explain the magnitude and the position of the observed feature. 
The anisotropy of   \textit{an individual} molecular orbital 
(in spite of the orbital-spin SU(4) and space $C_{4v}$ symmetries \cite{note2}) is the key for this splitting.
The hybridization with the conduction states broadens the VHS  but the splitting persists. 

\begin{figure}[tbp]
\includegraphics[scale=0.3, viewport= 80 34 700 550, clip]{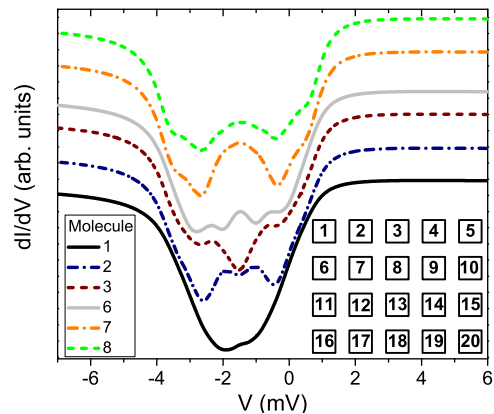} 
\caption{(Color online) differential conductance as a function of voltage
for several sites of a $5 \times 4$ cluster. The figures have been displaced vertically for clarity.
}
\label{cluster}
\end{figure}

The $dI/dV$ has been measured at different sites of a finite cluster, 
to study the effects of coordination on the observed spectra \cite{Tsukahara10_Evolution_of_Kondo_resonance}. In order to compare with experiment, 
we have applied our theory to a finite cluster of $5 \times 4$ molecules, 
as shown in \Fig{cluster} (see Ref. \onlinecite{note3}). Some of the curves 
display an oscillatory behavior, which are likely to disappear
for a more realistic calculation \cite{note3} or 
in the presence of 
disorder or inhomogeneities (not considered here). 
In any case, the results provide definite conclusions: the differential conductance at the corners
(sites of coordination number $Z=2$) do not show a splitting, while those with $Z=4$ do show two 
dips in the FKA. The sites with $Z=3$ display an intermediate and variable behavior 
which depends on the specific site.
These results also agree with the experimental trends \cite{Tsukahara10_Evolution_of_Kondo_resonance}. 

\textit{Summary and discussion.-} 
Motivated by recent experiments \cite{Gao07_Site_specific_Kondo_effect,Minamitani12_SU4_Kondo_in_FePc_molecules,Tsukahara10_Evolution_of_Kondo_resonance}, 
we have derived a Hubbard-Anderson model describing a square lattice of 
magnetic atoms or molecules with orbital degeneracy on top of a metallic surface. 
Extension to other lattices is straightforward. While the model has the $C_{4v}$ symmetry of the 
square lattice, the individual molecular orbitals are coupled via an anisotropic hopping which 
leads to two strongly renormalized VHS in the density of states of $3d$ electrons. 
The hybridization to the substrate broadens these VHS, but 
these features persist and dominate the density of states observed by the STM tip, 
therefore displaying two dips in the $dI/dV$ around $V=0$. 
We conclude that these VHS are the main explanation of the experimentally observed splitting in the FKA.
Our results explain the observed 
behavior in systems of FePc molecules
on Au(111), for an isolated molecule, the lattice,
and the evolution between them in a consistent way.

Our work has its own interest beyond FePc molecules.  
A study of a similar 2D model without coupling to the substrate,
suggests a ferromagnetic (FM) orbital ordering and antiferromagnetic (AFM) spin ordering at 
$T=0$ for small Hund's rule exchange \cite{Aligia04_Magnetic_and_orbital_order_in_ruthenates}.
The nearest-neighbor AFM interactions are of the order of 
$4 t_2^2/U \approx 10$ K or $4 t_1^2/U \approx 1$ K, depending on direction 
[see Eq. (10) of Ref. \onlinecite{Aligia04_Magnetic_and_orbital_order_in_ruthenates}], 
which are of the order of the Kondo temperature 
$T_K \approx 4.7$ K estimated from the half width at half maximum of the FKA. In addition, while the RKKY interaction $I$ 
is unlikely to explain the splitting of the FKA, it might also introduce interesting competing effects 
\cite{note} (also see Appendix).
For our specific system, a preliminary calculation based on the Stoner criterion shows that magnetic order 
would occur for $|I|> 16.1$ K.
While fluctuations in 2D destroy long-range magnetic order at  finite temperature, 
this opens the intriguing possibility of observing quantum critical behavior at 
low enough temperatures in 2D molecular Kondo systems. Indeed, the existence of orbitally-ordered phases 
\cite{Aligia04_Magnetic_and_orbital_order_in_ruthenates} and dissipative quantum phase transitions 
\cite{Lobos12_Dissipative_XY_chain, Lobos13_FMchains} have been suggested in related systems. 
Recently, long-range FM order was observed 
for a 2D layer of organic molecules absorbed on graphene
\cite{garnica13}. In transition-metal phtalocyanines the  coupling to the substrate is very sensitive to the particular 
transition-metal atom \cite{Mugarza12_Metal_Pc_on_Ag100, *Gargiani13_Metal_Pc_molecules_on_Au110}. We also expect a strong dependence on the substrate, as for example replacing
Au by Ag or Cu. Therefore new physics is likely to appear in the near future, and our theory (or
some modifications of it) is expected to bring valuable insight.

Acknowledgements. The authors are grateful to N. Takagi and E. Minamitani for useful comments. AML acknowledges support from JQI-NSF-PFC. 
MR and AAA thank CONICET from Argentina for financial support. This work was
partially supported by PIP 11220080101821 and PIP 11200621 of CONICET and
PICT R1776 of the ANPCyT, Argentina.

\begin{appendix}
\begin{widetext}
\section{Derivation of the effective model for a 2D lattice of FePc molecules on Au(111)}\label{app: effective_model}

To describe a 2D lattice of FePc molecules on Au(111), we need to estimate the effective hopping between relevant molecular orbitals of different molecules in the system. 
This hopping might take place indirectly through conduction states, as described in Appendix \ref{hop}, or through the molecular ligands 
of the molecules. To show the essential physics of the latter, we propose the simplified system of FeN$_{4}$ molecules 
(the central part of
FePc) shown in Fig. \ref{fig:lattice_simplified} in this supplemental material, and in Fig. 1(b) in the main manuscript. 
According to recent
\textit{ab-initio} calculations (see Ref. \cite{Minamitani12_SU4_Kondo_in_FePc_molecules})
the relevant orbitals in the $3d$-shell of the Fe atom are the degenerate
orbitals $d_{xz}$ and $d_{yz}$, depicted in blue and red in Fig.
\ref{fig:lattice_simplified}, respectively. The circles correspond to the N atoms. Only the projection of
the orbitals onto the $xy$-plane is shown in Fig. \ref{fig:lattice_simplified}.
In this model, the Fe atoms are connected via effective N-N links (i.e., dashed lines in Fig. \ref{fig:lattice_simplified}),
which encode the couplings via the benzene rings in the FePc molecule. Note that the Fe $3d_{\nu z}$ orbitals within a single 
FeN$_4$ substructure
hybridizes only with the $p_{z}$ orbitals of the N atoms in the $\nu $
direction ($\nu =x$ or $y$). The hopping of the $3d_{\nu z}$ orbital with
other $s$ or $p$ orbitals of N or in the other direction vanishes by
symmetry. 
We introduce the following basis of FeN$_4$
molecular states

\begin{align*}
\left|\tilde{x},1\right\rangle  & =\alpha\left|\tilde{d}_{xz},1\right\rangle +\beta\left[\left|\tilde{p}_z,4\right\rangle 
-\left|\tilde{p}_z,2\right\rangle \right],\\
\left|\tilde{y},1\right\rangle  & =\alpha\left|\tilde{d}_{yz},1\right\rangle +\beta\left[\left|\tilde{p}_z,1\right\rangle 
-\left|\tilde{p}_z,3\right\rangle \right],\\
\left|\tilde{x},2\right\rangle  & =\alpha\left|\tilde{d}_{xz},2\right\rangle +\beta\left[\left|\tilde{p}_z,8\right\rangle 
-\left|\tilde{p}_z,6\right\rangle \right],\\
\left|\tilde{y},2\right\rangle  & =\alpha\left|\tilde{d}_{yz},2\right\rangle +\beta\left[\left|\tilde{p}_z,5\right\rangle 
-\left|\tilde{p}_z,7\right\rangle \right],\\
\left|\tilde{x},3\right\rangle  & =\alpha\left|\tilde{d}_{xz},3\right\rangle +\beta\left[\left|\tilde{p}_z,12\right\rangle 
-\left|\tilde{p}_z,10\right\rangle \right],\\
\left|\tilde{y},3\right\rangle  & =\alpha\left|\tilde{d}_{yz},3\right\rangle +\beta\left[\left|\tilde{p}_z,9\right\rangle 
-\left|\tilde{p}_z,11\right\rangle \right],
\end{align*}
with the condition $|\alpha|^{2}+2|\beta|^{2}=1$, 
where the states $\left|\tilde{d}_{xz},i\right\rangle $ and $\left|\tilde{d}_{yz},i\right\rangle $
correspond to the Fe $d_{xz}$ and $d_{yz}$ orbitals in the $i$-th
molecule, and the states $\left|\tilde{p}_z,j\right\rangle $ correspond to the
$p_{z}$ orbitals at the $j$-th N atom (see Fig. \ref{fig:lattice_simplified}).

Using the tight-binding approximation
assuming a hopping $t^{\prime}$ between nearest-neigbor N atoms of
different FeN$_4$ molecules , we now compute the following matrix elements
between nearest-neighbor molecules:

\begin{align}
\left\langle \tilde{x},1\right|H_{mol}\left|\tilde{x},2\right\rangle  & =0, & \qquad & \left\langle \tilde{x},1\right|H_{mol}\left|\tilde{x},3\right\rangle =-t\nonumber \\
\left\langle \tilde{x},1\right|H_{mol}\left|\tilde{y},2\right\rangle  & =t, &  & \left\langle \tilde{x},1\right|H_{mol}\left|\tilde{y},3\right\rangle =-t,\nonumber\\
\left\langle \tilde{y},1\right|H_{mol}\left|\tilde{x},2\right\rangle  & =t, &  & \left\langle \tilde{y},1\right|H_{mol}\left|\tilde{y},3\right\rangle =0\nonumber \\
\left\langle \tilde{y},1\right|H_{mol}\left|\tilde{y},2\right\rangle  & =-t, &  & \left\langle \tilde{y},1\right|H_{mol}\left|\tilde{x},3\right\rangle =-t,\label{eq:matrix_elements} 
\end{align}
where $t=|\beta|^{2}t^{\prime}$. We note that although $\left\langle \tilde{x},i\right|H_{mol}\left|\tilde{y},j\right\rangle $
vanishes by symmetry for $i=j$, this is not the case for different
$i\neq j$. This fact in general complicates the theoretical description,
and we therefore introduce the unitary transformation on every site

\begin{align}
\left|x,i\right\rangle  & =\gamma\left|\tilde{x},i\right\rangle +\delta\left|\tilde{y},i\right\rangle ,\label{eq:xprime1}\\
\left|y,i\right\rangle  & =-\delta\left|\tilde{x},i\right\rangle +\gamma\left|\tilde{y},i\right\rangle ,\label{eq:yprime1}
\end{align}
 with the normalization condition $\delta=\sqrt{1-\gamma^{2}}$. So
far, the parameter $\gamma$ is arbitrary. The idea now is to look
for a particular basis $\left\{|x,i\rangle,|y,i\rangle \right\}$ with the property $\left\langle x,i\right|H_{mol}\left|y,j\right\rangle =0$
for all $i,j$. This is done by choosing a proper $\gamma$, and the procedure amounts to rotating anticlockwise the 
$\hat{x}$-$\hat{y}$ axes an angle $\theta=\arctan{\left[\sqrt{1-\gamma^2}/\gamma\right]}$. Since
in the same molecule and for next nearest (and more distant) neighbors,
the matrix element vanishes, we only have to focus on nearest neighbors,
e.g. $\left\langle x,1\right|H_{mol}\left|y,2\right\rangle =0$.
Using Eqs. \ref{eq:matrix_elements}, \ref{eq:xprime1} and \ref{eq:yprime1},
we obtain the equation 
\begin{align}
0  & =2\gamma^{2}-1-\gamma\sqrt{1-\gamma^{2}},\label{eq:gamma2_equation}
\end{align}
 %
whose solution is $\gamma^{2}=\frac{1}{2}\left(1+\sqrt{\frac{1}{5}}\right)\approx0.7236$
(the other solution of the quadratic equation corresponds to exchanging
$\gamma\longleftrightarrow\delta$). We next compute the new matrix
elements in the rotated basis between, e.g. the states $\left|x,1\right\rangle $
and $\left|x,2\right\rangle $, and $\left|y,1\right\rangle $
and $\left|y,2\right\rangle $ 

\begin{align}
\left\langle x,1\right|H_{mol}\left|x,2\right\rangle  
& =\left(\gamma\left\langle \tilde{x},1\right|+\sqrt{1-\gamma^{2}}\left\langle \tilde{y},1\right|\right)H_{mol}
\left(\gamma\left|\tilde{x},2\right\rangle +\sqrt{1-\gamma^{2}}\left|\tilde{y},2\right\rangle \right),\nonumber \\
 & =t\left[2\gamma\sqrt{1-\gamma^{2}}-\left(1-\gamma^{2}\right)\right],\nonumber \\
 & \approx0.6180t,\label{eq:matrix_element12_x_rotated}
\end{align}
 
\begin{align}
\left\langle y,1\right|H_{mol}\left|y,2\right\rangle  
& =\left(-\sqrt{1-\gamma^{2}}\left\langle \tilde{x},1\right|+\gamma\left\langle \tilde{y},1\right|\right)H_{mol}
\left(-\sqrt{1-\gamma^{2}}\left|\tilde{x},2\right\rangle +\gamma\left|\tilde{y},2\right\rangle \right),\nonumber \\
 & =t\left[-2\gamma\sqrt{1-\gamma^{2}}-\gamma^{2}\right],\nonumber \\
 & \approx-1.6180t.\label{eq:matrix12_element_y_rotated}
\end{align}

Similarly, between the states $\left|x,1\right\rangle $ and $\left|x,3\right\rangle $,
and $\left|y,1\right\rangle $ and $\left|y,3\right\rangle $, the matrix elements are  
$\left\langle x,1\right|H_{mol}\left|x,3\right\rangle   \approx-1.6180t$, and 
$\left\langle y,1\right|H_{mol}\left|y,3\right\rangle  \approx0.6180t,\label{eq:matrix_element_y_rotated-1}$

\begin{figure}
\includegraphics[viewport=20bp 50bp 400bp 450bp,clip,scale=0.5]{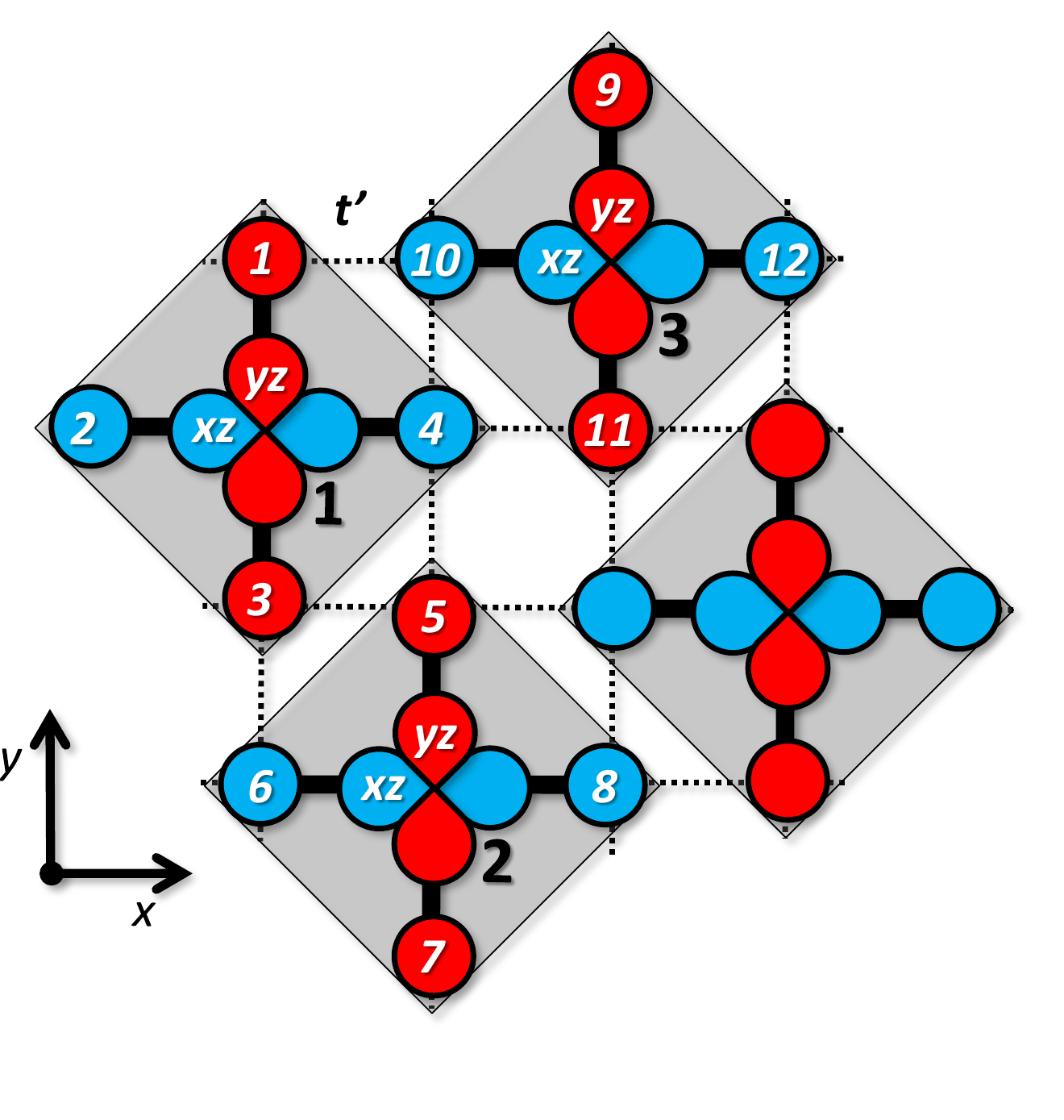}
\caption{Representation of the lattice of FePc molecules deposited on the top
of a Au(111) surface.\label{fig:lattice_simplified}}
\end{figure}

Therefore, although in this new basis the orbitals $\left|x,i\right\rangle $
and $\left|y,j\right\rangle $ are not coupled, note that
now\textit{ both} $\left|x,i\right\rangle $ and $\left|y,i\right\rangle $
disperse along the new axes $\hat{x}$ and $\hat{y}$.

In Appendix \ref{hop}, it is shown that the effective hopping through conduction states in the substrate shows the same features:
it is highly anisotropic and conserves the orbital index. 

Our goal now is to derive an effective model for the electrons that
occupy the new molecular orbitals. Based on Eqs. (\ref{eq:xprime1})
and (\ref{eq:yprime1}), we introduce the fermionic creation operators
$d_{\mathbf{r}_{ij},\sigma}^{x\dagger}$, $d_{\mathbf{r}_{ij},\sigma}^{y\dagger}$, which
create an electron on the transformed molecular orbitals $\left|x\right\rangle $,
$\left|y\right\rangle $ respectively, at site $\mathbf{r}_{ij}$
with spin $\sigma$ in the 2D Hubbard lattice of Fe sites. The effective
model then becomes

\begin{align}
H & =H_{mol}+H_{mix}+H_{c},\label{eq:H}\\
H_{mol} & =\sum_{ij}^N\sum_{\sigma}\left[E_{d}\left(d_{\mathbf{r}_{ij},\sigma}^{x\dagger}d_{\mathbf{r}_{ij},\sigma}^{x}
+d_{\mathbf{r}_{ij},\sigma}^{y\dagger}d_{\mathbf{r}_{ij},\sigma}^{y}\right)
-\left(t_{2}d_{\mathbf{r}_{ij},\sigma}^{x\dagger}d_{\mathbf{r}_{i+1,j},\sigma}^{x}
+t_{1}d_{\mathbf{r}_{ij},\sigma}^{x\dagger}d_{\mathbf{r}_{i,j+1},\sigma}^{x}+\text{H.c.}\right) \right.\nonumber\\
& -\left. \left(t_{1}d_{\mathbf{r}_{ij},\sigma}^{y\dagger}d_{\mathbf{r}_{i+1,j},\sigma}^{y}
+t_{2}d_{\mathbf{r}_{ij},\sigma}^{y\dagger}d_{\mathbf{r}_{i,j+1},\sigma}^{y}
+\text{H.c.}\right)\right]+\frac{U}{2}\sum_{ij}^N\left(\sum_{\sigma,\nu=\left\{x,y\right\}}  
d_{\mathbf{r}_{ij},\sigma}^{\nu\dagger}d_{\mathbf{r}_{ij},\sigma}^{\nu}  \right)
\left(\sum_{\sigma\nu}d_{\mathbf{r}_{ij},\sigma}^{\nu\dagger}d_{\mathbf{r}_{ij},\sigma}^{\nu}-1\right),\label{eq:Hd}\\
H_{mix} & =\frac{V}{\sqrt{M}}\sum_{\xi}\sum_{ij}^N\sum_{\sigma,\nu=\left\{x,y\right\}}
\left(d_{\mathbf{r}_{ij},\sigma}^{\nu\dagger}a_{\mathbf{r}_{ij},\xi,\sigma}^{\nu}
+\text{H.c.}\right),\label{eq:Hmix}\\
H_{c} & =\sum_{\xi}\sum_{\sigma,\nu=\left\{x,y\right\}}
\epsilon_{\xi}a_{\mathbf{r}_{ij},\xi,\sigma}^{\nu\dagger}a_{\mathbf{r}_{ij},\xi,\sigma}^{\nu},\label{eq:Hc}
\end{align}

where the hopping matrix elements are $t_{1}$ and $t_{2}$, with $|t_{1}|<|t_{2}|$.
The term $H_{mix}$ couples the orbitals $d_{xz}$ and $d_{yz}$ with
the metallic states in the substrate at the 2D position $\mathbf{r}_{ij}$ in the surface. 
The fermionic annihilation operators $a_{\mathbf{r}_{ij},\xi,\sigma}^{\nu}$
represent metallic conduction states which hybridize with the molecular state at $\mathbf{r}_{ij}$ with quantum numbers $\xi$.
More details of the hybridization are given in Appendix \ref{hop}. The effect of the conduction
states is equivalent to consider independent baths for each molecular state, and an effective
hopping between nearest molecules already included in $t_i$. 
This has been already found
in problems with a few sites using equations of motion \cite{Romero11_STM_for_adsorbed_molecules}.
Effective hoppings at larger distances are neglected. 
Based on this, we can describe $H_c$, the Hamiltonian describing the metal, as a collection of independent 
``local baths" at each Hubbard site $\mathbf{r}_{ij}$ \cite{Lobos12_Dissipative_XY_chain, Lobos13_FMchains}. 

The Hamiltonian (\ref{eq:H}) is SU(4)-invariant. To see this, one can show that the SU(4) rotations (i.e., exponentials of the SU(4) algebra generators) commute with $H$. For one site, in the basis $\left\vert x,\uparrow \right\rangle $, $\left\vert x,\downarrow
\right\rangle $,  $\left\vert y,\uparrow \right\rangle $, $\left\vert
y,\downarrow \right\rangle $, the SU(4) generators can be written as trivial diagonal matrices, permutations of two basis sets, or permutations
 with a change of phases for the permuted states \cite{Sbaih2013_SU4_generators}. For the lattice, similar generators can be constructed taking into account the additional translational symmetry of the square lattice. All nontrivial generators can be constructed from $C_{4}$ rotations for one spin only, permutations of spins for one orbital only, or products of three of these operations with a change of phases ($H$ remains invariant under this change, since it conserves orbital and spin indices).

Following Minamitami et al. \cite{Minamitani12_SU4_Kondo_in_FePc_molecules}, 
we assume the occupation of the degenerate orbitals $d_{xz}$
and $d_{yz}$ is between 3 and 4. Therefore, the Kondo effect is most
likely ocurring for a hole in our sysytem of molecular orbitales with $xz$ and $yz$ symmetry. 
We can simplify the
description of the problem introducing the electron-hole transformation

\begin{align}
d_{\mathbf{r}_{ij},\sigma}^{\nu} & \rightarrow\left(-1\right)^{i+j}h_{\mathbf{r}_{ij},\sigma}^{\nu\dagger},
\label{eq:eh_transf_d}\\
a_{\mathbf{r}_{ij},\xi,\sigma}^{\nu} & \rightarrow\left(-1\right)^{i+j+1}c_{\mathbf{r}_{ij},\xi,\sigma}^{\nu\dagger},
\label{eq:eh_transf_c}
\end{align}
and use a hole-representation of the electronic degrees of freedom,
which now represent fluctuations between states with $n=1$ and $n=0$ holes in
the molecular states. 
We now assume that the energies of configurations with $n>1$ holes
are much higher than those with $n=1$ and $n=0$. This can be effectively
expressed introducing the constrained slave-boson representation

\begin{align}
h_{\mathbf{r}_{ij},\sigma}^{\nu} & =b_{\mathbf{r}_{ij}}^{\dagger}f_{\mathbf{r}_{ij},\sigma}^{\nu},\label{eq:slave_boson_representation}\\
b_{\mathbf{r}_{ij}}^{\dagger}b_{\mathbf{r}_{ij}}+\sum_{\sigma,\nu=\left\{x,y\right\}}
f_{\mathbf{r}_{ij},\sigma}^{\nu\dagger}f_{\mathbf{r}_{ij},\sigma}^{\nu} & =1.\label{eq:constraint}
\end{align}
 The projected Hamiltonian for holes is therefore
\begin{align}
H_{h} & =\sum_{ij}^N\sum_{\sigma}\left[ \left(E_{h}+\lambda_{\mathbf{r}_{ij}}\right)
\left(f_{\mathbf{r}_{ij},\sigma}^{x\dagger}f_{\mathbf{r}_{ij},\sigma}^{x}+f_{\mathbf{r}_{ij},\sigma}^{y\dagger}
f_{\mathbf{r}_{ij},\sigma}^{y}\right)-\left(t_{2}b_{\mathbf{r}_{ij}}^{\dagger}
b_{\mathbf{r}_{i+1,j}}f_{\mathbf{r}_{ij},\sigma}^{x\dagger}f_{\mathbf{r}_{i+1,j},\sigma}^{x}
+t_{1}b_{\mathbf{r}_{ij}}^{\dagger}b_{\mathbf{r}_{i,j+1}}f_{\mathbf{r}_{ij},\sigma}^{x\dagger}f_{\mathbf{r}_{i,j+1},\sigma}^{x}
+\text{H.c.}\right) \right. \nonumber \\
 & \left.-\left(t_{1}b_{\mathbf{r}_{ij}}^{\dagger}b_{\mathbf{r}_{i+1,j}}f_{\mathbf{r}_{ij},\sigma}^{y\dagger}
f_{\mathbf{r}_{i+1,j},\sigma}^{y}+t_{2}b_{\mathbf{r}_{ij}}^{\dagger}b_{\mathbf{r}_{i,j+1}}
f_{\mathbf{r}_{ij},\sigma}^{y\dagger}f_{\mathbf{r}_{i,j+1},\sigma}^{y}+\text{H.c.}\right)\right]
+\sum_{i,j}^N\lambda_{ij}\left(b_{\mathbf{r}_{ij}}^{\dagger}b_{\mathbf{r}_{ij}}-1\right),\label{eq:Hh_sb}\\
H_{mix} & =\frac{V}{\sqrt{M}}\sum_{ij}^N b_{\mathbf{r}_{ij}}\sum_{\sigma,\nu=\left\{x,y\right\}}
\left(f_{\mathbf{r}_{ij},\sigma}^{\nu\dagger}c_{\mathbf{r}_{ij},\xi,\sigma}^{\nu}+\text{H.c.}\right),
\label{eq:Hmix_sb}\\
H_{c} & =\sum_{\xi}\sum_{\sigma,\nu=\left\{x,y\right\}}
\epsilon_{\xi}c_{\mathbf{r}_{ij},\xi,\sigma}^{\nu\dagger}c_{\mathbf{r}_{ij},\xi,\sigma}^{\nu},
\label{eq:Hc_lba}
\end{align}
 where $\lambda_{\mathbf{r}_{ij}}$ is a local Lagrange multiplier that inforces the contraint
Eq. (\ref{eq:constraint}) at site $\mathbf{r}_{ij}$. In Eq. (\ref{eq:Hh_sb}) we have defined
the diagonal energies for holes $E_{h}  \equiv -E_{d}-3U$,
 and we have neglected a constant contribution $4E_{d}+6U$ per site.

\subsection{Path integral formulation}

The partition function $Z$ of the system is given by the coherent
state functional integral 
\begin{align}
Z & =\int\mathcal{D}\left[\bar{f},f\right]\mathcal{D}\left[\bar{b},b\right]\mathcal{D}\left[\bar{c},c\right]d\lambda\; 
e^{-\mathcal{S}},\label{eq:Z}
\end{align}
 where $\mathcal{S}$ is the Euclidean action of the total system

\begin{align}
\mathcal{S} & =\int_{0}^{\beta}d\tau\;\left\{\sum_{i,j}^N 
\left[\sum_{\sigma,\nu}\bar{f}_{\mathbf{r}_{ij},\sigma}^{\nu}\left(\tau\right)
\left(\partial_{\tau}-\mu\right)f_{\mathbf{r}_{ij},\sigma}^{\nu}\left(\tau\right) 
+\bar{b}_{\mathbf{r}_{ij}}\left(\tau\right)\partial_{\tau}b_{\mathbf{r}_{ij}}\left(\tau\right) 
+\sum_{\xi,\sigma,\nu}\bar{c}_{\mathbf{r}_{ij},\xi,\sigma}^{\nu}\left(\tau\right)\left(\partial_{\tau}
-\mu\right)\bar{c}_{\mathbf{r}_{ij},\xi,\sigma}^{\nu}\left(\tau\right)\right] +H\left(\tau\right)\right\}.\label{eq:S}
\end{align}
Now, we perform the large-$\mathcal{N}$ approximation 
(where $\mathcal{N}=\mathcal{N}_{\text{spin}}\times\mathcal{N}_{\text{orbital}}$
are the total number of degenerate states in the $3d$ shell), which allows to perform
the semiclassical approximation for the bosonic variables \cite{coleman84, Coleman87, newns87}

\begin{align}
b_{\mathbf{r}_{ij}}\left(\tau\right) & \approx\left\langle b_{\mathbf{r}_{ij}}\left(\tau\right)\right\rangle 
=z_{\mathbf{r}_{ij}},\label{eq:sbmfa}
\end{align}
 where $z_{\mathbf{r}_{ij}}$ is a $c-$number representing the value of the condensed
boson. Assumming, in addition, translational invariance in the 2D
Hubbard lattice, which allows to set $z_{\mathbf{r}_{ij}}=z$ and $\lambda_{\mathbf{r}_{ij}}=\lambda$,
we obtain the action 
\begin{align}
\mathcal{S} & \approx\mathcal{S}_{f}^{0}+\mathcal{S}_{c}^{0}+\mathcal{S}_{mix}+\beta N\lambda\left(z^{2}-1\right),\label{eq:S_mf}
\end{align}
 with 
\begin{align}
\mathcal{S}_{f}^{0} & =\int_{0}^{\beta}d\tau\;\sum_{ij}^N\left[\sum_{\sigma,\nu}
\bar{f}_{\mathbf{r}_{ij},\sigma}^{\nu}\left(\tau\right)\left(\partial_{\tau}+E_{h}
+\lambda-\mu\right)f_{\mathbf{r}_{ij},\sigma}^{\nu}\left(\tau\right)
-\left(t_{2}z^{2}\bar{f}_{\mathbf{r}_{ij},\sigma}^{x}\left(\tau\right)
f_{\mathbf{r}_{i+1,j},\sigma}^{x}\left(\tau\right)+t_{1}z^{2}\bar{f}_{\mathbf{r}_{ij},\sigma}^{x}
\left(\tau\right)f_{\mathbf{r}_{i,j+1},\sigma}^{x}\left(\tau\right)+\text{H.c.}\right)\right.\nonumber \\
 & \left.-\left(t_{1}z^{2}\bar{f}_{\mathbf{r}_{ij},\sigma}^{y}\left(\tau\right)
f_{\mathbf{r}_{i+1,j},\sigma}^{y}\left(\tau\right)+t_{2}z^{2}\bar{f}_{\mathbf{r}_{ij},\sigma}^{y}
\left(\tau\right)f_{\mathbf{r}_{i,j+1},\sigma}^{y}\left(\tau\right)+\text{H.c.}\right)\right],\label{eq:S0_f}\\
\mathcal{S}_{mix} & =\int_{0}^{\beta}d\tau\;\sum_{ij}^N\sum_{\sigma\nu}
\frac{Vz}{\sqrt{M}}\bar{f}_{\mathbf{r}_{ij},\sigma}^{\nu}\left(\tau\right)c_{\mathbf{r}_{ij},\xi,\sigma}^{\nu}
\left(\tau\right)+\text{H.c.},\label{eq:S_mix}\\
\mathcal{S}_{c}^{0} & =\int_{0}^{\beta}d\tau\;\sum_{i,j}^N\sum_{\xi,\sigma,\nu}\bar{c}_{\mathbf{r}_{ij},\xi,\sigma}^{\nu}
\left(\tau\right)\left(\partial_{\tau}+\epsilon_{\xi}-\mu\right)c_{\mathbf{r}_{ij},\xi,\sigma}^{\nu}\left(\tau\right),
\label{eq:S0_c}
\end{align}
and where the values of $z$ and $\lambda$ are obtained from the
stepest-descent method (see below). In Fourier representation

\begin{align}
f_{\mathbf{r}_{ij},\sigma}^{\nu}\left(\tau\right) & =\frac{1}{\sqrt{\beta N}}
\sum_{\mathbf{k},\omega_{n}}e^{i\mathbf{k}.\mathbf{r}_{ij}-i\omega_{n}\tau}
f_{\mathbf{k},\sigma}^{\nu}\left(i\omega_{n}\right),\label{eq:ft_f}\\
c_{\mathbf{r}_{ij},\xi,\sigma}^{\nu}\left(\tau\right) & =\frac{1}{\sqrt{\beta N}}
\sum_{\mathbf{k},\omega_{n}}e^{i\mathbf{k}.\mathbf{r}_{ij}-i\omega_{n}\tau}c_{\mathbf{k},\xi,\sigma}^{\nu}
\left(i\omega_{n}\right),\label{eq:ft_c}
\end{align}
 where $\mathbf{k}$ is the 2D momentum in the plane and $\omega_{n}=\frac{2\pi}{\beta}\left(n+\frac{1}{2}\right)$
are the fermionic Matsubara frequencies. In this representation, the
action becomes 
\begin{align}
\mathcal{S}_{f}^{0} & =\sum_{\mathbf{k},\omega_{n}}\sum_{\sigma,\nu}
\bar{f}_{\mathbf{k},\sigma}^{\nu}\left(i\omega_{n}\right)\left[-i\omega_{n}+E_{h}
+\lambda-\mu-2z^{2}\left(t_{2}\cos k_{x}a_{0}+t_{1}\cos k_{y}a_{0}\right)\right]
f_{\mathbf{k},\sigma}^{\nu}\left(i\omega_{n}\right),\label{eq:S0_f_ft}\\
\mathcal{S}_{mix} & =\sum_{\mathbf{k},\omega_{n}}\sum_{\sigma\nu}\frac{Vz}{\sqrt{M}}
\left[\bar{f}_{\mathbf{k},\sigma}^{\nu}\left(i\omega_{n}\right)
c_{\mathbf{k},\xi,\sigma}^{\nu}\left(i\omega_{n}\right)
+\bar{c}_{\mathbf{k},\xi,\sigma}^{\nu}\left(i\omega_{n}\right)
f_{\mathbf{k},\sigma}^{\nu}\left(i\omega_{n}\right)\right],\label{eq:Smix_ft}\\
\mathcal{S}_{c}^{0} & =\sum_{\mathbf{k},\omega_{n}}\sum_{\xi,\sigma,\nu}\bar{c}_{\mathbf{k},\xi,\sigma}^{\nu}
\left(i\omega_{n}\right)\left[-i\omega_{n}+\epsilon_{\xi}-\mu\right]c_{\mathbf{k},\xi,\sigma}^{\nu}
\left(i\omega_{n}\right),\label{eq:S0_c_ft}
\end{align}
where in Eq. (\ref{eq:S0_f_ft}) we have used the degeneracy of the
two bands $\nu=\left\{x,y\right\}$. It is convenient to rewrite the partition function
after these manipulations 
\begin{align}
Z & =e^{-\beta N\lambda\left(z^{2}-1\right)}\int\mathcal{D}\left[\bar{f},f\right]\mathcal{D}\left[\bar{c},c\right]\; 
e^{-\mathcal{S}_{f}^{0}-\mathcal{S}_{c}^{0}-\mathcal{S}_{mix}}.\label{eq:Z_sbmfa}
\end{align}
We now focus on the 2D system and integrate out the fermionic degrees
of freedom in the conduction band 
\begin{align}
Z & =e^{-\beta N\lambda\left(z^{2}-1\right)}Z_{c}^{0}\frac{\int\mathcal{D}\left[\bar{c},c\right]\; e^{-\mathcal{S}_{c}^{0}}
\int\mathcal{D}\left[\bar{f},f\right]\; e^{-\mathcal{S}_{f}^{0}-\mathcal{S}_{mix}}}{Z_{c}^{0}},\nonumber \\
 & =e^{-\beta N\lambda\left(z^{2}-1\right)}Z_{c}^{0}\int\mathcal{D}\left[\bar{f},f\right]\; 
e^{-\mathcal{S}_{eff}\left[\bar{f},f\right]},\label{eq:Z_sbmfa_2}
\end{align}
where we have defined the effective action for $f$ electrons 
\begin{align}
\mathcal{S}_{eff}\left[\bar{f},f\right] & =\sum_{\mathbf{k},\omega_{n}}
\sum_{\sigma,\nu}\bar{f}_{\mathbf{k},\sigma}^{\nu}\left(i\omega_{n}\right)
\left[-i\omega_{n}+E_{h}+\lambda-\mu-2z^{2}\left(t_{2}\cos k_{x}a_{0}
+t_{1}\cos k_{y}a_{0}\right)+\frac{V^{2}z^{2}}{M}\sum_{\xi}G_{\mathbf{k},\xi,\nu,\sigma}^{cc0}
\left(i\omega_{n}\right)\right]f_{\mathbf{k},\sigma}^{\nu}\left(i\omega_{n}\right),\label{eq:S_eff}
\end{align}
and
{} 
\begin{align}
G_{\mathbf{k},\xi,\nu,\sigma}^{cc0}\left(i\omega_{n}\right) & =-\int_{0}^{\beta}d\tau\; e^{i\omega_{n}\tau}\left\langle 
T_{\tau}c_{\mathbf{k},\xi,\sigma}^{\nu}\left(\tau\right)\bar{c}_{\mathbf{k},\xi,\sigma}^{\nu}\left(0\right)\right\rangle ,
\nonumber \\
 & =\frac{1}{i\omega_{n}-\left(\epsilon_{q}-\mu\right)}.\label{eq:Gaa0}
\end{align}
 is the unperturbed Matsubara Green's function for conduction electrons. The same result as in Eq. \ref{eq:S_eff} above can be obtained with the mean-field Hamiltonian
\begin{align}
H_{\text{MF}} & =\sum_{\mathbf{k},\sigma,\nu}\left(E_{h}+\lambda-\mu-2t_{2}z^{2}\cos k_{x}a_{0}-2t_{1}z^{2}\cos k_{y}a_{0}\right)
f_{\mathbf{k},\sigma}^{\nu\dagger}f_{\mathbf{k},\sigma}^{\nu}\nonumber \\
 & +\frac{Vz}{\sqrt{M}}\sum_{\mathbf{k},\xi,\sigma,\nu}\left(f_{\mathbf{k},
\sigma}^{\nu\dagger}c_{\mathbf{k},\xi,\sigma}^{\nu}+\text{H.c.}\right)+\sum_{\mathbf{k},\xi,\sigma,\nu}
\left(\epsilon_{\xi}-\mu\right)c_{\mathbf{k},\xi,\sigma}^{\nu\dagger}c_{\mathbf{k},\xi,\sigma}^{\nu}.\label{eq:H_meanfield}
\end{align}

\subsection{Calculation of the free-energy and minimization}

The free-energy of the 2D Hubbard system in Eq. (\ref{eq:S_eff}) or Eq. (\ref{eq:H_meanfield}) is calculated using standard
techniques as

\begin{align}
\Delta F_{f} & =-\frac{1}{\beta}\ln\left(Z/Z_{c}^{0}\right)\nonumber \\
 & =-\frac{1}{\beta}\ln\left[e^{-\beta N\lambda\left(z^{2}-1\right)}\int\mathcal{D}\left[\bar{f},f\right]\; 
e^{-\mathcal{S}_{eff}\left[\bar{f},f\right]}\right],\nonumber \\
 & =N\lambda\left(z^{2}-1\right)-\frac{1}{\beta}\sum_{\mathbf{k},\omega_{n}}\sum_{\sigma,\nu}\ln\left[-i\omega_{n}+E_{h}+
\lambda-2z^{2}\left(t_{2}\cos k_{x}a_{0}+t_{1}\cos k_{y}a_{0}\right)+\frac{V^{2}z^{2}}{M}\sum_{\xi}G_{\mathbf{k},\xi,
\nu,\sigma}^{cc0}\left(i\omega_{n}\right)\right],\label{eq:F}
\end{align}
where we have set the chemical potential to $\mu=0$. At this point, we assume the conduction band to be flat and featureless 
at the Fermi
energy, which allows to perform the approximation

\begin{align}
\frac{1}{M}\sum_{\xi}G_{\mathbf{k},\xi,\nu,\sigma}^{cc0}\left(i\omega_{n}\right) 
& =G_{\nu,\sigma}^{cc0}\left(i\omega_{n}\right)\label{eq:Gcc0_local}\\
 & =\frac{1}{M}\sum_{\xi}\frac{1}{i\omega_{n}-\epsilon_{\xi}},\\
 & =-i\rho_{0}\ln\left(\frac{i\omega_{n}-W}{i\omega_{n}+W}\right),\label{eq:Gcc0_local_explicit}
\end{align}
where $W$ is half of the bandwidth in the conduction band and $\rho_{0}=1/2W$
is the density of states per spin and channel. After this approximation,
and using the $SU\left(\mathcal{N}\right)$ symmetry of the problem
(at the end we will set $\mathcal{N}=4$), we finally obtain the free-energy
per site 
\begin{align}
\Delta f & =\lambda\left(z^{2}-1\right)-\frac{\mathcal{N}}{\beta N}\sum_{\mathbf{k},\omega_{n}}\ln\left[-i\omega_{n}
+E_{h}+\lambda-2z^{2}\left(t_{2}\cos k_{x}a_{0}+t_{1}\cos k_{y}a_{0}\right)-iz^{2}V^{2}\rho_{0}\ln\left(\frac{i\omega_{n}
-W}{i\omega_{n}+W}\right)\right].\label{eq:f}
\end{align}
 The Matsubara sum is evaluated as

\begin{align}
\Delta f & =\lambda\left(z^{2}-1\right)+\frac{\mathcal{N}}{N}\sum_{\mathbf{k}}\frac{1}{\pi}\int_{-\infty}^{\infty}d\omega\; 
n_{F}\left(\omega\right)\text{Im}\ln\left[-\omega+E_{h}+\lambda-2z^{2}\left(t_{2}\cos k_{x}a_{0}+t_{1}\cos k_{y}a_{0}\right)
-iz^{2}V^{2}\rho_{0}\pi\theta\left(W-\left|\omega\right|\right)\right],\nonumber \\
 & =\lambda\left(z^{2}-1\right)-\frac{\mathcal{N}}{N}\sum_{\mathbf{k}}\frac{1}{\pi}\int_{-W}^{W}d\omega\; n_{F}\left(\omega\right)\arctan\left[\frac{z^{2}\Gamma}{-\omega+E_{h}+\lambda-2z^{2}\left(t_{2}\cos k_{x}a_{0}+t_{1}\cos k_{y}a_{0}\right)}\right],\label{eq:f2}
\end{align}
 where we have defined the inverse lifetime of the $f-$electrons,
$\Gamma\equiv\pi\rho_{0}V^{2}$. At $T=0$, the free-energy coincides
with the groundstate energy and the expression simplifies to%

\begin{align}
\Delta e_{g} & =\lambda\left(z^{2}-1\right)-\frac{\mathcal{N}}{N}\sum_{\mathbf{k}}\frac{1}{\pi}\int_{-W}^{0}d\omega\;\text{ArcTan}\left[\frac{z^{2}\Gamma}{-\omega+E_{h}+\lambda-2z^{2}\left(t_{2}\cos k_{x}a_{0}+t_{1}\cos k_{y}a_{0}\right)}\right],\nonumber \\
 & =\lambda\left(z^{2}-1\right)-\frac{\mathcal{N}}{\pi}z^{2}\Gamma-\frac{\mathcal{N}}{\pi^{2}}\int_{0}^{\pi/a}dk_{x}
\int_{0}^{\pi/a}dk_{y}\Biggl\{\left[E_{h}+\lambda-2z^{2}\left(t_{2}\cos k_{x}a_{0}+t_{1}\cos k_{y}a_{0}\right)\right]\nonumber \\
 & \times\left[\frac{1}{\pi}\text{ArcTan}\left(\frac{E_{h}+\lambda-2z^{2}\left(t_{2}\cos k_{x}a_{0}+t_{1}\cos k_{y}a_{0}\right)}
{z^{2}\Gamma}\right)-\frac{1}{2}\right]\\
 & \left.-\frac{z^{2}\Gamma}{2\pi}\ln\left[\frac{\left(E_{h}+\lambda-2z^{2}\left(t_{2}\cos k_{x}a_{0}+t_{1}\cos k_{y}a_{0}\right)\right)^{2}+\left(z^{2}\Gamma\right)^{2}}{W^{2}}\right]\right\} \label{eq:eg}
\end{align}
 Introducing the change of variables 
\begin{eqnarray}
x & = & -\cos\left(k_{x}a_{0}\right),\nonumber \\
dx & = & \sin\left(k_{x}a_{0}\right)d\left(k_{x}a_{0}\right)=\sqrt{1-x^{2}}d\left(k_{x}a_{0}\right),\label{eq:ch_variable_x}
\end{eqnarray}
 
\begin{eqnarray}
y & = & -\cos\left(k_{y}a_{0}\right),\nonumber \\
dy & = & \sin\left(k_{y}a_{0}\right)d\left(k_{y}a_{0}\right)=\sqrt{1-y^{2}}d\left(k_{y}a_{0}\right),\label{eq:ch_variable_y}
\end{eqnarray}
 we can write the integral as 
\begin{align}
\Delta e_{g} & =\lambda\left(z^{2}-1\right)-\frac{\mathcal{N}}{\pi}z^{2}\Gamma-\frac{\mathcal{N}}{\pi^{2}}\int_{-1}^{1}\frac{dx}
{\sqrt{1-x^{2}}}\int_{-1}^{1}\frac{dy}{\sqrt{1-y^{2}}}\nonumber \\
 & \times\Biggl\{\left[E_{h}+\lambda+2z^{2}\left(t_{2}x+t_{1}y\right)\right]\left[\frac{1}{\pi}\text{ArcTan}\left(\frac{E_{h}
+\lambda+2z^{2}\left(t_{2}x+t_{1}y\right)}{z^{2}\Gamma}\right)-\frac{1}{2}\right]\\
 & -\frac{z^{2}\Gamma}{2\pi}\ln\left[\frac{\left(E_{h}+\lambda+2z^{2}\left(t_{2}x+t_{1}y\right)\right)^{2}+\left(z^{2}
\Gamma\right)^{2}}{W^{2}}\right]\Biggr\}.\label{eq:eg2}
\end{align}
 The minimum of the energy is found by minimization with respect to
$\lambda$ and $z^{2}$ %
\begin{align}
\frac{\partial\Delta e_{g}}{\partial\lambda}=0 & =\left(z^{2}-1\right)
-\frac{\mathcal{N}}{\pi^{2}}\int_{-1}^{1}\frac{dx}{\sqrt{1-x^{2}}}\int_{-1}^{1}\frac{dy}{\sqrt{1-y^{2}}}
\left[\frac{1}{\pi}\text{ArcTan}\left(\frac{E_{h}+\lambda+2z^{2}t_{2}x+2z^{2}t_{1}y}{z^{2}\Gamma}\right)-\frac{1}{2}\right],
\label{eq:selfconsistent_lambda}\\
\frac{\partial\Delta e_{g}}{\partial z^{2}}=0 & =\lambda-\frac{\mathcal{N}}{\pi^{2}}\int_{-1}^{1}
\frac{dx}{\sqrt{1-x^{2}}}\int_{-1}^{1}\frac{dy}{\sqrt{1-y^{2}}}\Biggl\{2\left(t_{2}x+t_{1}y\right)
\left[\frac{1}{\pi}\text{ArcTan}\left(\frac{E_{h}+\lambda+2z^{2}t_{2}x+2z^{2}t_{1}y}{z^{2}\Gamma}\right)
-\frac{1}{2}\right]\nonumber \\
 & -\frac{\Gamma}{2\pi}\ln\left[\frac{\left(E_{h}+\lambda+2z^{2}t_{2}x+2z^{2}t_{1}y\right)^{2}
+\left(z^{2}\Gamma\right)^{2}}{W^{2}}\right]\Biggr\},\label{eq:selfconsistent_z}
\end{align}

\subsection{\label{sec:isolated_impurity}Isolated impurity limit}

We return to Eqs. (\ref{eq:selfconsistent_lambda}) and (\ref{eq:selfconsistent_z})
and set the parameters $t_{1}=t_{2}=0$. We then recover the limit
of the isolated impurity:

\begin{align}
\frac{\partial\Delta e_{g}}{\partial\lambda}=0 & =\left(z^{2}-1\right)
-\mathcal{N}\left[\frac{1}{\pi}\text{ArcTan}\left(\frac{E_{h}+\lambda}{z^{2}\Gamma}\right)-\frac{1}{2}\right],\label{eq:selfconsistent_lambda_impurity}\\
\frac{\partial\Delta e_{g}}{\partial z^{2}}=0 & =\lambda+\frac{\mathcal{N}\Gamma}{2\pi}\ln\left[\frac{\left(E_{h}
+\lambda\right)^{2}+\left(z^{2}\Gamma\right)^{2}}{W^{2}}\right],\label{eq:selfconsistent_z_impurity}
\end{align}
 We obtain the parameter $z$ from Eq. (\ref{eq:selfconsistent_z_impurity})
\begin{align}
z^{2} & =\sqrt{\frac{W^{2}e^{-\frac{2\pi\lambda}{\mathcal{N}\Gamma}}-\left(E_{h}+\lambda\right)^{2}}{\Gamma^{2}}},
\end{align}
 which yields the equation for $\lambda$ 
\begin{align}
0 & =1-\sqrt{\frac{W^{2}e^{-\frac{2\pi\lambda}{\mathcal{N}\Gamma}}-\left(E_{h}
+\lambda\right)^{2}}{\Gamma^{2}}}+\mathcal{N}\left[\frac{1}{\pi}\text{ArcTan}\left(\frac{E_{h}
+\lambda}{\sqrt{W^{2}e^{-\frac{2\pi\lambda}{\mathcal{N}\Gamma}}-\left(E_{h}+\lambda\right)^{2}}}\right)
-\frac{1}{2}\right].
\end{align}

\subsection{Calculation of the local density of $d-$states}

Once the the solutions $\left(\lambda_{0},z_{0}^{2}\right)$ of Eqs.
(\ref{eq:selfconsistent_lambda}) and (\ref{eq:selfconsistent_z})
are obtained for a particular set of parameters $E_{h},\Gamma,W$,
and $r=t_{2}/t_{1}$ of the model, we can calculate different quantities
of interest. Here we focus on the local density of $d-$states 
\begin{align}
\rho_{d}\left(\omega\right) & =-\frac{1}{\pi}\text{Im }G_{\mathbf{r}_{ij}}^{dd}
\left(i\omega_{n}\rightarrow\omega+i0^{+}\right),\label{eq:rho_d}\\
 & =-\frac{1}{\pi}\sum_{\mathbf{k},\nu,\sigma}\text{Im }G_{\mathbf{k},\nu,\sigma}^{dd}
\left(i\omega_{n}\rightarrow\omega+i0^{+}\right),\label{eq:rho_d_sum_k}
\end{align}
 where 
\begin{align}
G_{\mathbf{k},\nu,\sigma}^{dd}\left(i\omega_{n}\right) & =-\int_{0}^{\beta}d\tau\; 
e^{-i\omega_{n}\tau}\left\langle T_{\tau}d_{\mathbf{k},\sigma}^{\nu}
\left(\tau\right)d_{\mathbf{k},\sigma}^{\nu\dagger}\left(0\right)\right\rangle .\label{eq:Gdd}
\end{align}
 However, the results from the previous sections are in the language
of \textit{holes} (see Eqs. (\ref{eq:eh_transf_d}) and (\ref{eq:eh_transf_c})).
Therefore, the idea now is to relate our knowledge of the quantities
in this language, to the quantities of interest in the electron language,
using the transformations (\ref{eq:eh_transf_d}) ``backwards''.
Using the general relation $\left\langle T_{\tau}A\left(\tau\right)B\left(0\right)\right\rangle 
=\eta\left\langle T_{\tau}B\left(-\tau\right)A\left(0\right)\right\rangle $
($\eta=+1$ for bosons, $\eta=-1$ for fermions), we obtain the relation
between Green's functions:%

\begin{align}
G_{\mathbf{k},\nu,\sigma}^{dd}\left(i\omega_{n}\right) & =-G_{\mathbf{k},\nu,\sigma}^{hh}\left(-i\omega_{n}\right).\label{eq:Gdd_Ghh_relation}
\end{align}
 This relation is valid in general. In the SBMF approximation, $G_{\mathbf{k},\nu,\sigma}^{dd}\left(i\omega_{n}\right)
\approx-z^{2}G_{\mathbf{k},\nu,\sigma}^{ff}\left(-i\omega_{n}\right)$,
and therefore we obtain the explicit form 
\begin{align}
G_{\mathbf{k},\nu,\sigma}^{dd}\left(i\omega_{n}\right) & =\frac{z^{2}}{i\omega_{n}+E_{h}+\lambda-2z^{2}\left(t_{2}\cos k_{x}a+t_{1}\cos k_{y}a\right)+iz^{2}\Gamma\theta\left(W-\left|\omega\right|\right)},\label{eq:Gdd_sbmfa}
\end{align}
 where Eqs. (\ref{eq:S_eff}) and (\ref{eq:Gcc0_local_explicit})
have been used. We now replace this expression into Eq. (\ref{eq:rho_d_sum_k})
to obtain the LDOS of $d$-states. We introduce the change of variables
in Eqs. (\ref{eq:ch_variable_x}) and (\ref{eq:ch_variable_y}) to
compute the double integral over momentum, and obtain 
\begin{align}
G_{\mathbf{r}_{ij}}^{dd}\left(\omega\right) & =z^{2}\mathcal{N}\left(\frac{1}{2\pi}\right)^{2}
\int dk_{x}dk_{y}\frac{1}{\omega+E_{h}+\lambda-2z^{2}\left(t_{2}\cos k_{x}a+t_{1}\cos k_{y}a\right)
+iz^{2}\Gamma\theta\left(W-\left|\omega\right|\right)},\nonumber \\
 & =\frac{z^{2}\mathcal{N}}{\pi^{2}}\int_{-1}^{1}dx\int_{-1}^{1}dy\frac{1}{\sqrt{1-x^{2}}}
\frac{1}{\sqrt{1-y^{2}}}\frac{1}{\omega+E_{h}+\lambda-2z^{2}\left(t_{2}x+t_{1}y\right)
+iz^{2}\Gamma\theta\left(W-\left|\omega\right|\right)}.\label{eq:Gdd_local}
\end{align}

\subsection{Calculation of the local density of conduction states}

In the STM experiment, the observed quantity is the differential conductance
$dI/dV$ as a function of the gate voltage $V$ (the voltage between
the STM tip and the substrate). $dI/dV$ is directly proportional
to the density of a mixed operator, which contains information of
the conduction electrons and the localized electrons weghted by its
hopping to the tip of the STM \cite{al05}:

\begin{align}
\rho_{t}\left(\omega\right) & =-\frac{1}{\pi}\sum_{\sigma,\nu}\text{Im }\left[G_{\mathbf{r}_{ij},\nu,\sigma}^{tt}\left(\omega+i0^{+}\right)\right].\label{eq:rho_t_conduction}
\end{align}
Introducing the following notation for the Matsubara Green's functions \cite{zubarev60,mahan}, we can express
\begin{align}
G_{\mathbf{r}_{ij},\nu,\sigma}^{tt}\left(i\omega_{n}\right) & =\left\langle \left\langle t_{\mathbf{r}_{ij},\sigma}^{\nu};
t_{\mathbf{r}_{ij},\sigma}^{\nu\dagger}\right\rangle \right\rangle _{i\omega_{n}}\equiv-\int_{0}^{\beta}d\tau\; 
e^{-i\omega_{n}\tau}\left\langle T_{\tau}t_{\mathbf{r}_{ij},\sigma}^{\nu}\left(\tau\right)
t_{\mathbf{r}_{ij},\sigma}^{\nu\dagger}\left(0\right)\right\rangle ,\label{eq:Gtt}\\
G_{\mathbf{r}_{ij},\nu,\sigma}^{dd}\left(i\omega_{n}\right) & =\left\langle \left\langle 
d_{\mathbf{r}_{ij},\sigma}^{\nu};d_{\mathbf{r}_{ij},\sigma}^{\nu\dagger}\right\rangle 
\right\rangle _{i\omega_{n}}\equiv-\int_{0}^{\beta}d\tau\; e^{-i\omega_{n}\tau}\left\langle 
T_{\tau}d_{\mathbf{r}_{ij},\sigma}^{\nu}\left(\tau\right)d_{\mathbf{r}_{ij},\sigma}^{\nu\dagger}\left(0\right)\right\rangle ,
\label{eq:Gdd-1}\\
G_{\mathbf{r}_{ij},\xi,\nu,\sigma}^{cc}\left(i\omega_{n}\right) & =\left\langle 
\left\langle c_{\mathbf{r}_{ij},\xi,\sigma}^{\nu}; 
c_{\mathbf{r}_{ij},\xi,\sigma}^{\nu\dagger}\right\rangle \right\rangle _{i\omega_{n}}\equiv-\int_{0}^{\beta}d\tau\; e^{-i\omega_{n}\tau}\left\langle T_{\tau}c_{\mathbf{r}_{ij},\xi,\sigma}^{\nu}\left(\tau\right)c_{\mathbf{r}_{ij},\xi,\sigma}^{\nu\dagger}\left(0\right)\right\rangle ,\label{eq:Gcc-1}\\
G_{\mathbf{r}_{ij},\xi,\nu,\sigma}^{cd}\left(i\omega_{n}\right) & =\left\langle \left\langle 
c_{\mathbf{r}_{ij},\xi,\sigma}^{\nu};d_{\mathbf{r}_{ij},\sigma}^{\nu\dagger}\right\rangle 
\right\rangle _{i\omega_{n}}\equiv-\int_{0}^{\beta}d\tau\; e^{-i\omega_{n}\tau}\left\langle 
T_{\tau}c_{\mathbf{r}_{ij},\xi,\sigma}^{\nu}\left(\tau\right)d_{\mathbf{r}_{ij},\sigma}^{\nu\dagger}\left(0\right)\right\rangle .\label{eq:Gcd}
\end{align}
where we have defined the operator
\begin{align}
t_{\mathbf{r}_{ij},\sigma}^{\nu} & \equiv\frac{1}{\sqrt{M}}
\sum_{\xi}c_{\mathbf{r}_{ij},\xi,\sigma}^{\nu}+qd_{\mathbf{r}_{ij},\sigma}^{\nu},\label{eq:t}
\end{align}
i.e., the linear combination of conduction and localized holes seen by the STM tip at site $\mathbf{r}_{ij}$. Replacing this definition into
Eq. (\ref{eq:Gtt}), we can express $G_{\mathbf{r}_{ij},\nu,\sigma}^{tt}\left(i\omega_{n}\right)$
in terms of the other Green's functions: 
\begin{align*}
G_{\mathbf{r}_{ij},\nu,\sigma}^{tt}\left(i\omega_{n}\right) & =\frac{1}{M}\sum_{\xi, \xi^{\prime}}
G_{\mathbf{r}_{ij},\xi,\xi^{\prime},\nu,\sigma}^{cc}\left(i\omega_{n}\right)
+\frac{q}{\sqrt{M}}\sum_{\xi}\left[G_{\mathbf{r}_{ij},\xi,\nu,\sigma}^{cd}\left(i\omega_{n}\right)
+G_{\mathbf{r}_{ij},\xi,\nu,\sigma}^{dc}\left(i\omega_{n}\right)\right]+q^{2}G_{\mathbf{r}_{ij},\nu,\sigma}^{dd}\left(i\omega_{n}\right),\\
 & =\frac{1}{N}\sum_{\mathbf{k}}\left[\frac{1}{M}\sum_{\xi, \xi^{\prime}}
G_{\mathbf{k},\xi, \xi^{\prime},\nu,\sigma}^{cc}\left(i\omega_{n}\right)+\frac{q}{\sqrt{M}}
\sum_{\xi}\left(G_{\mathbf{k},\xi,\nu,\sigma}^{cd}\left(i\omega_{n}\right)
+G_{\mathbf{k},\xi,\nu,\sigma}^{dc}\left(i\omega_{n}\right)\right)\right.\\
 & \left.+q^{2}G_{\mathbf{k},\nu,\sigma}^{dd}\left(i\omega_{n}\right)\right],
\end{align*}
 where Eqs. (\ref{eq:ft_f}) and (\ref{eq:ft_c}) have been used.
Using equations of motion (see Appendix \ref{sec:eqn_motion}), we
can express all the Green's function of the problem in terms of $G_{\mathbf{k},\nu,\sigma}^{dd}\left(i\omega_{n}\right)$
and $G_{\mathbf{k},q,\nu,\sigma}^{0cc}\left(i\omega_{n}\right)$.
We obtain the result:
\begin{align}
G_{\mathbf{r}_{ij},\nu,\sigma}^{tt}\left(i\omega_{n}\right) & =\frac{1}{N}\sum_{\mathbf{k}}
\frac{1}{M}\sum_{\xi,\xi^{\prime}}\left[\delta_{\xi,\xi^{\prime}}
G_{\mathbf{k},\xi,\nu,\sigma}^{0cc}\left(i\omega_{n}\right)+\frac{V^{2}}{M}
G_{\mathbf{k},\xi,\nu,\sigma}^{0cc}\left(i\omega_{n}\right)
G_{\mathbf{k},\xi^{\prime},\nu,\sigma}^{0cc}\left(i\omega_{n}\right)
G_{\mathbf{k},\nu,\sigma}^{dd}\left(i\omega_{n}\right)\right]+\nonumber \\
 & +\frac{1}{N}\sum_{\mathbf{k}}\frac{1}{M}\sum_{\xi}\left[2qVG_{\mathbf{k},\xi,\nu,\sigma}^{0cc}\left(i\omega_{n}\right)
G_{\mathbf{k},\nu,\sigma}^{dd}\left(i\omega_{n}\right)\right]+\frac{1}{N}
\sum_{\mathbf{k}}\left[q^{2}G_{\mathbf{k},\nu,\sigma}^{dd}\left(i\omega_{n}\right)\right],\nonumber \\
 & =G_{\nu,\sigma}^{0cc}\left(i\omega_{n}\right)
+\left[VG_{\nu,\sigma}^{cc0}\left(i\omega_{n}\right)+q\right]^{2}\left[\frac{1}{N}\sum_{\mathbf{k}}
G_{\mathbf{k},\nu,\sigma}^{dd}\left(i\omega_{n}\right)\right],\label{eq:Gtt_final}
\end{align}
where Eqs. (\ref{eq:Gcc0_local}), (\ref{eq:Gcc}) and (\ref{eq:Gdc}) have been used. We now return to 
the expression for the local DOS Eq. (\ref{eq:rho_t_conduction}).
Substracting the background of conduction electrons (first term in
the above Eq. (\ref{eq:Gtt_final})) and replacing by the expressions
(\ref{eq:Gcc0_local_explicit}) and (\ref{eq:Gdd_sbmfa}), we finally
obtain

\begin{align}
\Delta\rho_{t}\left(\omega\right) & =-\frac{1}{\pi}\sum_{\sigma,\nu}
\text{Im }\left[G_{\mathbf{r}_{ij},\nu,\sigma}^{tt}\left(\omega+i0^{+}\right)
-G_{\nu,\sigma}^{0cc}\left(\omega+i0^{+}\right)\right],\\
 & =-\frac{1}{\pi}\text{Im }\left[\left(-i\rho_{0}V\ln\left(\frac{\omega+i0^{+}-W}{\omega+i0^{+}+W}\right)
+q\right)^{2}\right.\\
 & \left.\times\left(\frac{z^{2}\mathcal{N}}{\pi^{2}}
\int_{-1}^{1}dx\int_{-1}^{1}dy\frac{1}{\sqrt{1-x^{2}}}\frac{1}{\sqrt{1-y^{2}}}\frac{1}{\omega+E_{h}
+\lambda-2z^{2}\left(t_{2}x+t_{1}y\right)+iz^{2}\Gamma\theta\left(W-\left|\omega\right|\right)}\right)\right]\label{eq:diff_rho_surface}
\end{align}

\subsection{\label{sec:eqn_motion}Equations of motion}

The following are general relations, valid independently of any approximation. 
We follow the methods and definitions in Ref. \onlinecite{zubarev60}. 
We start from the equation of motion for $G_{\mathbf{k},\xi, \xi^{\prime},\nu,\sigma}^{cc}\left(i\omega_{n}\right)$:
\begin{align}
i\omega_{n}\left\langle \left\langle c_{\mathbf{k},\xi,\sigma}^{\nu};
c_{\mathbf{k},\xi^{\prime},\sigma}^{\nu\dagger}\right\rangle \right\rangle _{i\omega_{n}} 
& =\delta_{\xi,\xi^{\prime}}+\left\langle \left\langle \left[c_{\mathbf{k},\xi,\sigma}^{\nu},H\right];
c_{\mathbf{k},\xi^{\prime},\sigma}^{\nu\dagger}\right\rangle \right\rangle _{i\omega_{n}}\nonumber \\
 & =\delta_{\xi,\xi^{\prime}}+\left(\epsilon_{q}-\mu\right)\left\langle 
\left\langle c_{\mathbf{k},\xi,\sigma}^{\nu};
c_{\mathbf{k},\xi^{\prime},\sigma}^{\nu\dagger}\right\rangle \right\rangle _{i\omega_{n}}
+\frac{V}{\sqrt{M}}\left\langle \left\langle d_{\mathbf{k},\sigma}^{\nu};
c_{\mathbf{k},\xi^{\prime},\sigma}^{\nu\dagger}\right\rangle \right\rangle _{i\omega_{n}},\label{eq:}
\end{align}
where the Hamiltonian $H$ is defined in Eq. (\ref{eq:H}). From here we obtain the expression 
\begin{align}
G_{\mathbf{k},\xi,\xi^{\prime},\nu,\sigma}^{cc}\left(i\omega_{n}\right) 
& =\delta_{\xi,\xi^{\prime}}G_{\mathbf{k},\xi,\nu,\sigma}^{0cc}\left(i\omega_{n}\right)
+\frac{V}{\sqrt{M}}G_{\mathbf{k},\xi,\nu,\sigma}^{0cc}\left(i\omega_{n}\right)G_{\mathbf{k},\xi^{\prime},\nu,\sigma}^{dc}\left(i\omega_{n}\right),\label{eq:Gcc}
\end{align}
where we have used the definition Eq. (\ref{eq:Gaa0}). 
Then we compute the equation of motion for $G_{\mathbf{k},\xi^{\prime},\nu,\sigma}^{dc}\left(i\omega_{n}\right)$:
\begin{align}
i\omega_{n}\left\langle \left\langle d_{\mathbf{k},\sigma}^{\nu};
c_{\mathbf{k},\xi^{\prime},\sigma}^{\nu\dagger}\right\rangle \right\rangle _{i\omega_{n}} 
& =\left\langle \left\langle d_{\mathbf{k},\sigma}^{\nu};
\left[H,c_{\mathbf{k},\xi^{\prime},\sigma}^{\nu\dagger}\right]\right\rangle \right\rangle _{i\omega_{n}}\nonumber \\
 & =\left(\epsilon_{\xi^{\prime}}-\mu\right)\left\langle \left\langle d_{\mathbf{k},\sigma}^{\nu};
c_{\mathbf{k},\xi^{\prime},\sigma}^{\nu\dagger}\right\rangle \right\rangle _{i\omega_{n}}
+\frac{V}{\sqrt{M}}\left\langle \left\langle d_{\mathbf{k},\sigma}^{\nu};
d_{\mathbf{k},\sigma}^{\nu\dagger}\right\rangle \right\rangle _{i\omega_{n}}\label{eq:-1}
\end{align}
 From here we obtain
\begin{align}
G_{\mathbf{k},\xi^{\prime},\nu,\sigma}^{dc}\left(i\omega_{n}\right) 
& =\frac{V}{\sqrt{M}}G_{\mathbf{k},\xi^{\prime},\nu,\sigma}^{0cc}\left(i\omega_{n}\right)
G_{\mathbf{k},\nu,\sigma}^{dd}\left(i\omega_{n}\right).\label{eq:Gdc}
\end{align}
In this form, all the Green's fuctions of the problem are expressed
in terms of $G_{\mathbf{k},\xi,\nu,\sigma}^{0cc}\left(i\omega_{n}\right)$
and $G_{\mathbf{k},\nu,\sigma}^{dd}\left(i\omega_{n}\right)$.

\subsection{Effective hopping between molecular orbitals through the conduction
band}

\label{hop}

In this section we estimate the hybridization of the molecular orbitals
with the conduction band and  the effective hopping between molecular
orbitals through the conduction band. The hybridization of a conduction
state with wave vector $\mathbf{k}$\ described by a plane wave 
$\exp (i \mathbf{k\cdot r})$ and the molecular orbital with symmetry $z\nu $  located
at $\mathbf{R}$ is

\begin{equation}
V_{\nu }(\mathbf{R},\mathbf{k})=\int d\mathbf{r}\psi _{z\nu }^{\ast }(\mathbf{r-R})V(%
\mathbf{r-R})\exp (i\mathbf{k\cdot r}),  \label{v1}
\end{equation}%
where $\psi _{z\nu }(\mathbf{r})$ is the wave function of the molecular
orbital (with main weight on the corresponding $3d$ orbital) and $V(\mathbf{r%
})$ the potential. 

To make a simple estimate, we take the two dimensional case. Using polar
coordinates $\mathbf{k}=(k,\phi )$, $\mathbf{r-R}=(\rho ,\varphi )$, and
assuming  $V(\mathbf{r-R})\sim e^{2}/\rho $ one obtains

\begin{eqnarray}
V_{\nu }(\mathbf{R},k,\phi ) &=&\exp (i\mathbf{k\cdot R})V_{\nu }(\mathbf{0}%
,k,\phi ),  \notag \\
V_{x}(\mathbf{0},k,\phi ) &\sim &\int d\rho \Psi (\rho )\int d\varphi \cos
\varphi \exp [ik\rho \cos (\varphi -\phi )],  \notag \\
V_{y}(\mathbf{0},k,\phi ) &\sim &\int d\rho \Psi (\rho )\int d\varphi \sin
\varphi \exp [ik\rho \cos (\varphi -\phi )],  \label{v2}
\end{eqnarray}%
where $\Psi (\rho )$ is the radial part of $\psi _{z\nu }(\mathbf{r})$.
Assuming that $\Psi (\rho )$ is strongly localized within a distance $a$, so
that the Ferrmi wave vector $k_{F}\ll 1/a$, one can expand the exponential
up to first order in $k\rho $, and evaluate the angular integral. The radial
integral is of the order of $ka^{2}$. Then one has

\begin{equation}
V_{\nu }(\mathbf{0},\mathbf{k})=ak_{\nu }V_{0},  \label{v3}
\end{equation}%
where $k_{x}=k\cos \phi $, $k_{y}=k\sin \phi $ and $V_{0}$ is an energy.

From equations of motions (cf. Ref. \onlinecite{Romero11_STM_for_adsorbed_molecules}) or perturbation theory, 
one obtains that the
effective dynamical hopping between mollecular orbital $z\nu $ at position $%
\mathbf{0}$ and  $z\mu $ at $\mathbf{R}$ is

\begin{equation}
t_{\nu \mu }(\mathbf{R})=\sum_{\mathbf{k}}\frac{V_{\nu }^{\ast }(\mathbf{0},%
\mathbf{k})V_{\mu }(\mathbf{R},\mathbf{k})}{\omega -\epsilon _{\mathbf{k}}}.
\label{tnumu}
\end{equation}%
Evaluating this for $\omega $ on the Fermi shell, averaging over the
possible directions of $\mathbf{k}$ with $|\mathbf{k}|=k_{F}$  and assuming
for simplicity a symmetrical band with constant density of conduction states 
$\rho $, one obtains

\begin{eqnarray}
t_{\nu \mu }(\mathbf{R}) &=&\frac{F}{2\pi }\int d\phi \frac{k_{\nu }k_{\mu }%
}{k_{F}^{2}}\exp [ik_{F}R\cos (\theta -\phi )],  \notag \\
F &=&-i\pi \rho (ak_{F}V_{0})^{2},  \label{tnm}
\end{eqnarray}%
where in polar coordinates $\mathbf{R}=(R,\theta ).$

It is easy to see that symmetry imposes $t_{\nu \mu }\sim \delta _{\nu \mu }$. 
Also for $R=0$, the angular average is 1/2 and $t_{xx}(\mathbf{0})=t_{yy}(%
\mathbf{0})=-i\Gamma $ reduce to the self energy correction of an isolated
impurity due to hybridization with the conduction electrons. Thus $%
F=-2i\Gamma $. Evaluating the angular integral for $\theta =0$, we obtain
the two effective hoppings non equivalent by symmetry. They are

\begin{eqnarray}
t_{2} &=&t_{xx}(R,0)=-2i\Gamma \left[ \frac{J_{1}(k_{F}R)}{k_{F}R}%
-J_{2}(k_{F}R)\right] ,  \notag \\
t_{1} &=&t_{yy}(R,0)=-2i\Gamma \frac{J_{1}(k_{F}R)}{k_{F}R},  \label{tfinal}
\end{eqnarray}%
where $J_{n}(x)$ is the $n$-th Bessel function of the first kind.

For  $k_{F}R\approx 17.8$, more appropriate if bulk states dominate the hybridization 
($k_{F}=1.21$ /\AA , $R=14.7$ \AA)  
this gives $t_{2}/t_{1}=3.9$. Using the value  $\Gamma \simeq 10$ meV that
we obtained from our fits, the above estimate gives $|t_{2}|=0.81$ meV.
However, due to the oscillations of the Bessel functions, the values are 
very sensitive to the value of $k_{F}R$. For example for $k_{F}R=16$, one has  
$t_{2}/t_{1}=-17.7$ and $|t_{2}|=3.06$ meV. In any case, the anisotropy of the hopping is high.
If instead $k_{F}R=2.54$  is used (where we have used the value $k_F=0.173$/\AA \ corresponding to Shockley surface states 
\citep{Kevan87_Photoemission_in_111_noble_metals,Hyldgaard00_Long_range_adsorbate_interaction_in_111_metals}),
then $t_{2}/t_{1}=1.36$ and $|t_{2}|=5.20$ meV. It might be posible that both
bulk and surface states are important with the former dominating the width of
the resonance [$t_{\nu \mu }(\mathbf{0})$] and the latter dominating the
effective hybridization at larger distance.
We remark that for molecules arranged in a square lattice and using conservation of 
the orbital index, the phases of both $t_i$ can be gauged away and can take
both of them real and positive.

The estimated values of $|t_{2}|$ are smaller than that obtained from our fit
(21 meV) but in the same order of magnitude.

\section{Estimation of the Ruderman-Kittel-Kasuya-Yosida (RKKY) interaction}\label{app: rkky}

In Ref. \onlinecite{Tsukahara10_Evolution_of_Kondo_resonance} it is  suggested that the RKKY interaction might explain the observed
splitting of the Fano-Kondo resonance. Here we address in detail the estimation of 
the RKKY interaction for the case of 3D and 2D Au conduction states. Based on the results of this section, we conclude that the RKKY interaction is much smaller than the observed splitting $\Delta=4z^2||t_2|^2-|t_1|^2|\approx 29$ K, and is unlikely to explain it. Moreover, in the more realistic scenario where the magnetic impurities hybridize predominantly with bulk conduction electrons, the RKKY interaction is one order of magnitude smaller than $T_K$.

An advantage of Au and its (111) surface is that both the bulk sates
and the surface Shockley states near the Fermi energy can be described
as free electrons and therefore the calculations in the books by Kittel
\citep{kittel} for 3D and Ref. \onlinecite{beal87} for the 2D case
are valid. Following these works one can write for dimension N=2 or
3 for two spins $\mathbf{S}_{1}$ and $\mathbf{S}_{2}$ at a distance
$R$ (later we will consider nearest neighbors only)

\begin{equation}
H\left(R\right)=I_{\text{ND}}\left(R\right)\mathbf{S}_{1}.\mathbf{S}_{2},\label{h}
\end{equation}
with

\begin{equation}
I_{\text{ND}}\left(R\right)=-\frac{1}{4}\tilde{J}_{\text{ND}}^{2}\chi_{\text{ND}}\left(R\right).\label{i}
\end{equation}
where $\tilde{J}_{\text{3D}}=JV_{\text{at}}$ , $\tilde{J}_{\text{2D}}=JS_{\text{at}}$, 
$V_{\text{at}}$ ($S_{\text{at}}$) is the volume (surface) per
Au atom in the bulk (surface) and $\chi_{\text{ND}}\left(R\right)$
is the spin susceptiblity,  given by Eqs. (14) and (15) of Ref.
\onlinecite{beal87}.

\subsection{RKKY in the 3D case\label{sub:3D}}

The spin susceptibility of the 3D gas is \citep{beal87}

\begin{equation}
\chi_{\text{3D}}\left(R\right)=-\rho_{\text{3D}}\left(\epsilon_{F}\right)
\frac{4k_{F}^{3}}{\pi}F_{\text{3D}}\left(2k_{F}R\right),\label{s3}
\end{equation}
where $\rho_{\text{3D}}\left(\epsilon_{F}\right)=mk_{F}/2\pi^{2}\hbar^{2}$
is the density of states per spin and per unit volume, and

\begin{equation}
F_{\text{3D}}\left(x\right)\equiv\frac{x\cos-\sin x}{x^{4}}.\label{f3}
\end{equation}
Using the value $k_{F}=1.21$\AA \ $^{-1}$\citep{ashcroft}, one obtains
 $\rho_{\text{3D}}\left(\epsilon_{F}\right)=0.00805/\left(\text{eV.}\text{\AA}^{3}\right)$.
The density per atom and spin projection is $\rho=\rho_{\text{3D}}\left(\epsilon_{F}\right)V_{\text{at}}=0.137/$eV,
where we have used $V_{\text{at}}=17.0$ \AA{}$^{3}$ (the lattice
parameter of f.c.c. Au is $a=4.08$ \AA{}\ and $V_{\text{at}}=a^{3}/4$).
Imposing the condition $\tilde{J}_{\text{3D}}\rho_{\text{3D}}\left(\epsilon_{F}\right)=J\rho$,
in order to reproduce the observed $T_{K}\approx 5$ K, one obtains $J=0.8$ eV.

Using the above equations with $|F_{\text{3D}}\left(x\right)|\leq1/x^{3}$
for large $x$ and 
$R=14.7$ \AA{}\ as reported in Ref. \onlinecite{Tsukahara10_Evolution_of_Kondo_resonance} 
for the intermolecular distance we obtain%

\begin{equation}
|I_{\text{3D}}|\leq\left(JV_{\text{at}}\right)^{2}\rho_{\text{3D}}\left(\epsilon_{F}\right)\frac{1}{\pi(2R)^{3}}
=0.21\text{ K}\label{i3}
\end{equation}

This is roughly two orders of magnitude smaller than the splitting $\Delta$ and one order of magnitude smaller than $T_K$.

\subsection{RKKY in the 2D case\label{sub:2D}}

For the sake of completeness, it is instructive to study the effect of the 2D Schokley states in the RKKY 
interaction. In that case, the spin susceptibility for the 2D case is given by the expression
(14) of Ref. \onlinecite{beal87}

\begin{equation}
\chi_{\text{2D}}\left(R\right)=-\rho_{\text{2D}}\left(\epsilon_{F}\right)k_{F}^{2}F_{\text{2D}}\left(k_{F}R\right),\label{s2}
\end{equation}
where and $\rho_{\text{2D}}\left(\epsilon_{F}\right)=m^{\ast}/2\pi\hbar^{2}$,
and (note the absence of factor 2 in the argument of the function
$F_{\text{2D}}$) 
\begin{align}
F_{\text{2D}}\left(x\right) & \equiv J_{0}\left(x\right)Y_{0}\left(x\right)+J_{1}\left(x\right)Y_{1}\left(x\right),\notag\\
 & \xrightarrow[x\rightarrow\infty]{}-\frac{\sin\left(2x\right)}{\pi x^{2}}\label{f2}
\end{align}
with $J_{\nu}\left(x\right)$($Y_{\nu}\left(x\right)$) the Bessel
function of the first (second) kind \citep{abramowitz}. The effective
mass of the surface Shockley states is $m^{\ast}=0.28m$ \citep{Kevan87_Photoemission_in_111_noble_metals,Hyldgaard00_Long_range_adsorbate_interaction_in_111_metals}.
This leads to
$\rho_{\text{2D}}\left(\epsilon_{F}\right)=0.00589/\left(\text{eV}\AA^{2}\right)$.
Using $S_{\text{at}}=\sqrt{3}a^{2}/4=7.21\ \text{\AA}^{2}$ one obtains
$\rho=\rho_{\text{2D}}\left(\epsilon_{F}\right)S_{\text{at}}=0.0425/\text{eV}$.
Knorr \textit{et al.} have shown that the bulk states dominate the
hybridization with the impurity and therefore are more important than
the surface states in the Kondo screening \citep{knorr02}. Assuming
(as an overestimation) that half of the contribution to $J\rho$ is
due to surface states (and the other half to bulk states) leads to
$J=1.29$ eV. From Eq. (\ref{f2}) for large $x$, $|F_{\text{2D}}\left(x\right)|\leq1/\left(\pi x^{2}\right)$,
and using Eqs. (\ref{i}) and (\ref{s2}) for 
$R=14.7$ \AA{}\ we obtain

\begin{equation}
|I_{\text{2D}}|\leq\left(JS_{\text{at}}\right)^{2}\rho_{\text{2D}}\left(\epsilon_{F}\right)\frac{1}{4\pi R^{2}}
=2.17\text{ K.}\label{i2}
\end{equation}
The ground state energy of the model per site is -$|I_{\text{2D}}|/4$
for the ferromagnetic case and $-0.67I_{\text{2D}}$ for the antiferromagnetic
case \citep{Sauerwein94_AFM_Heisenberg_in_2D}. 
The above  result must be considered as an upper limit for the effect of 2D Schokley states 
on the RKKY interaction in FePc molecules on Au(111). While the magnitude of $I_{\text{2D}}$ is closer 
to the observed $T_K$, it is still much smaller than the observed splitting $\Delta$. 
Nevertheless, a more detailed study of the competition between the Kondo effect and the RKKY for the 
2D case in this class of  systems is interesting in its own right, and might have interesting implications 
for the quantum phase diagram due to the proximity to a quantum critical point \cite{doniach77a}. A preliminary result based on the Stoner criterion yields a critical value $|I_c|\approx 16.1$ K.


\end{widetext}
\end{appendix}

\bibliographystyle{apsrev}
\def\urlprefix{}
\def\url#1{}

\begin{thebibliography}{46}
\bibitem[{not({\natexlab{a}})}]{note}
\bibinfo{note}{In Ref. \onlinecite{Tsukahara10_Evolution_of_Kondo_resonance} it
  is suggested that the RKKY interaction $I$ might explain the observed
  splitting $\Delta$, but our estimations show that it is too small. Assuming
  that bulk Au electrons dominate the hybridization with the impurity
  \cite{knorr02,aaa06}, from the observed $T_K= W$ exp$[-1/(\rho J)] \approx 5$
  K and the reported $\rho=$ 0.137/eV one estimates a Kondo coupling $J=0.8$
  eV. Using the RKKY expression for free electrons and known data we obtain for
  the reported intermolecular distance $R = 14.7$ \AA, $|I_\text{3D}| \leq
  0.21$ K. Similarly, using known data for the 2D Schockley states and the
  expression for $I$ in 2D \cite{beal87}, assuming that half of $T_K$ the Kondo
  temperature is due to surface states, we obtain $|I_\text{2D}| \leq 2.17$ K.
  In any case $I < T_K \ll \Delta$ (see Appendix for more details).}
\bibitem[{not({\natexlab{d}})}]{notevhs}
\bibinfo{note}{A tight binding model in a rectangular lattice with hoppings
  $t_x$, $t_y$ in the $x,y$ directions has a dispersion relation $E(k_x,k_y)=-2
  (t_x \cos k_x + t_y \cos k_y)$. The sign of $t_x$ ($t_y$) can be changed by a
  gauge transformation $d_{i,j} \rightarrow (-1)^i d_{i,j}$ ($d_{i,j}
  \rightarrow (-1)^j d_{i,j}$). For $t_x,t_y >0$, the van Hove singularities
  are at the saddle points of $E(k_x,k_y)$ with wave vectors $(0,\pi)$ and
  $(\pi,0)$ and energies $E=2(t_y-t_x)$ and $E=2(t_x-t_y)$ respectively}.
\bibitem[{not({\natexlab{e}})}]{note2}
\bibinfo{note}{The actual point group in the experimental array of molecules
  \cite{Tsukahara10_Evolution_of_Kondo_resonance} is $C_4$ because the lines of
  molecules are not parallel to the main directions of the substrate. However,
  since we are modeling the substrate states with a continuum, the symmetry of
  our model is higher}.
\bibitem[{not({\natexlab{f}})}]{note3}
\bibinfo{note}{Since the minimization of the free energy in the SBMFA with
  respect to all $z_{\mathbf{r}_{ij}}$ and Lagrange multipliers
  $\lambda_{\mathbf{r}_{ij}}$ enforcing the constraints for all non-equivalent
  sites $\mathbf{r}_{ij}$ is a formidable task, we have made the additional
  assumption that $z_{\mathbf{r}_{ij}}=z$ and
  $\lambda_{\mathbf{r}_{ij}}=\lambda$, where $z$ and $\lambda$ are the
  corresponding values for the lattice. The dependence of these quantities on
  site would likely act in a similar way as disorder, blurring the oscillations
  observed in \Fig{cluster},}.
\end{thebibliography}

\end{document}